\documentclass[11pt]{article}
\usepackage{graphicx}
\usepackage{amsmath}
\usepackage{amssymb}
\usepackage{tikz}
\usepackage{color}
\usepackage{fullpage}
\begin{document}
	\title{An exponential integrator for the drift-kinetic model}
	\author{Nicolas Crouseilles,  Lukas Einkemmer, Martina Prugger}                     %
	\date{Received: date / Revised version: date}
	
	\maketitle
	
	\begin{abstract}
We propose an exponential integrator for the drift-kinetic equation in cylindrical geometry. 
This approach removes the CFL condition from the linear part of the system 
(which is often the most stringent requirement in practice) and treats the remainder explicitly 
using Arakawa's finite difference scheme. 
The present approach is mass conservative, 
up to machine precision, and significantly reduces the computational effort per time step. 

In addition, we demonstrate the efficiency of our method by performing numerical simulations 
in the context of the ion temperature gradient instability. In particular, we find that our numerical 
method can take time steps comparable to what has been reported in the literature for the (predominantly used) splitting 
approach. In addition, the proposed numerical method has significant advantages with respect to 
conservation of energy and efficient higher order 
methods can be obtained easily. We demonstrate this by investigating the performance of a fourth order implementation.
	\end{abstract}

	\section{Introduction}
	\label{intro}
	The evolution of strongly magnetized plasmas, like those encountered in tokamak devices, are governed by gyrokinetic type equations, 
	where the highly oscillatory motion of the charged particles around the magnetic field lines is averaged out. In a simplified slab geometry, gyrokinetic models
	reduce to the drift-kinetic equation, where the unknown $f$ depends on three cylindrical spatial coordinates and one velocity direction. This model is composed 
	of a guiding-center type dynamics in the plane orthogonal to the magnetic field lines and of a Vlasov type dynamics in the parallel (to the magnetic field lines) 
	direction. Moreover, the electric field is determined in a nonlinear way through a quasi-neutrality equation. 
	This drift-kinetic model is helpful to investigate plasma turbulence with an acceptable computational effort, and 
	consequently has received a lot of attention in recent decades.
	
	The goal of this work  is to develop accurate and efficient numerical methods for the drift-kinetic model.
	Several different numerical approaches have been proposed for the drift-kinetic equation or more generally for  kinetic equations. 
	Among them is the family of particle in cell (PIC) schemes which only discretizes the self-consistent electric and magnetic fields. Particles are, on the other hand, traced in phase space
	along the characteristics of the corresponding kinetic equation. 
	Although these methods are relatively cheap from a computational point of view, they suffer from inaccurate results in 
	low density regions and slow convergence as the number of macro-particles is increased (see, for example, \cite{pic, lee, lin}).
	
	As an alternative, Eulerian methods, such as finite volumes or finite differences, have been introduced. Since they perform a discretization 
	of the full phase space they are able to overcome many of the difficulties inherent in PIC schemes. 
	It should be duly noted, however, that such schemes are expensive (due to the high dimensionality of the problem) and the time step size is often limited by a stringent CFL condition. 
	The latter deficiency can be overcome by employing semi-Lagrangian schemes. 
	These methods use characteristic curves
	(to remove the stability constraint) but still maintain a discretization of the full phase space. 
	In a sense, semi-Lagrangian methods can be considered 
	a good compromise between the Lagrangian approach (for example, PIC) and the Eulerian approach. 
	Nevertheless, such methods often require large computational resources. Consequently, it is vital to develop more 
	efficient numerical methods to reduce the computational cost associated with Eulerian or semi-Lagrangian schemes.
	
	Within the semi-Lagrangian approach, splitting methods are very popular. 
	The reason for this is that (as has been realized in the seminal paper by Cheng \& Knorr \cite{cheng1976integration}) 
	splitting schemes often decouple different parts of the equation and allow us 
	to perform computations on lower-dimensional slices of phase space. 
	A significant part of the literature has been devoted to the construction 
	and analysis of splitting methods (see, for example, \cite{casas2016high,cheng1976integration,einkemmer2014convergence,einkemmer2014convergence2,klimas1994splitting}). 
	In addition, a range of space discretization schemes has been proposed. 
	Among these interpolation by cubic splines is perhaps the most popular (see, for example, \cite{klimas1994splitting,sonnendrucker1999semi}). Nevertheless, spectral methods 
	and methods based on discontinuous Galerkin type approximations have been considered as well (see, \cite{einkemmer2017Vlasov,rossmanith2011positivity}).
	
	In particular, for the drift-kinetic equation, which we consider in this paper, splitting methods have been employed extensively and production-level computer codes are available that implement such schemes (see \cite{gysela, grandgirard, latu2}). 
	However, while for the Vlasov--Poisson equation splitting allows us to reduce the problem to a sequence of one-dimensional advections, this is not true in the present case. In addition, we have to perform a three term splitting compared to only two terms in the Vlasov--Poisson equation. 
	Some high-order numerical methods that avoid splitting have also been proposed for gyrokinetic type models. The time integration is mostly based on explicit Runge-Kutta methods 
	(see \cite{candy, jenko}) which suffer from a CFL condition.

	In the present paper we propose an alternative to what has been considered in the literature. 
	In particular, we will use the observation that in many problems the most stringent 
	CFL condition is associated with the linear part of the drift-kinetic equation. 
	We propose to solve this linear part using Fourier techniques 
	as part of an exponential integrator. Note that high-order exponential integrators are easily available (see, for example, \cite{ei}). 
	A somewhat related idea has been used in \cite{idomura} for a more complex gyrokinetic model. There the linear 
	part (which involves the stiffest terms) is solved implicitly whereas the nonlinear part is treated explicitly. 
	In \cite{latu2}, the linear part is solved using a semi-Lagrangian approach where the function value at the feet of the characteristics is computed using a $4D$ interpolation. 
	In our approach, we exploit the structure of the drift-kinetic model by solving  the linear part exactly. 
	This can be done efficiently by using Fourier techniques. It should also be noted that contrary to some of the methods described above, our proposed scheme requires no multi-dimensional interpolation.

	The rest of this work is structured as follows. First, the drift-kinetic model is 
	introduced in section \ref{sec:dk}. Then, in section \ref{sec:method}, our proposed numerical 
	method is described and its (theoretical) properties are compared to that of the splitting approach. In section \ref{sec:numerical}, 
	we perform numerical simulations that allow us to conduct an extensive comparison of our exponential integrator with a splitting scheme. 
	Finally, we conclude in section \ref{sec:conclusion}.
	
	\subsection{Drift-kinetic equations \label{sec:dk}}
	In this work, we focus on the numerical approximation of the drift-kinetic model. 
	More specifically, our goal is to compute an approximation of $f=f(t,r,\theta,z,v)$ \color{black}using the initial condition $f_0(r,\theta,z,v)=f(t=0,r,\theta,z,v)$ \color{black}and satisfying the following $4D$ slab 
	drift-kinetic equation (for more details see \cite{vlasovia, grandgirard})
	\begin{equation}
	\label{dk}
	\partial_tf-\frac{\partial_\theta \phi}{r}\partial_rf+\frac{\partial_r \phi}{r}\partial_\theta f
	+v\partial_zf-\partial_z\phi\partial_{v}f=0,
	\end{equation}
    for $(r,\theta,z,v)\in \Omega\times[0,L]\times \mathbb{R}$, $\Omega=[r_{\rm min},r_{\rm max}]\times [0, 2\pi]$\color{black}, where $r_{max}>r_{min}>0$\color{black}.
	The self-consistent potential $\phi=\phi(r,\theta,z)$ is determined from the quasi-neutrality equation
	\begin{align}
	\label{qn}
	-\left[\partial_{r}^{2}\phi+\left(\frac{1}{r}+\frac{\partial_{r}n_{0}(r)}{n_{0}(r)}\right)\partial_{r}\phi+\frac{1}{r^{2}}\partial_{\theta}^{2}\phi\right]+\frac{1}{T_{e}(r)}\left(\phi-\left\langle \phi\right\rangle \right) \nonumber \\
	=\frac{1}{n_0(r)}\int_{\mathbb{R}} rfdv-1,
	\end{align}
	with $\langle\phi\rangle = \frac{1}{L}\int_0^L \phi(r,\theta,z)\,dz$. 
	For the potential, periodic boundary conditions 
	are considered in the $\theta$ and $z$ directions whereas in the radial direction, homogeneous Dirichlet boundary conditions are imposed.
	See \cite{vlasovia}, \cite{latu2} and \cite{dio} for more details.
	For the solution $f$, periodic boundary conditions are imposed in the $\theta, z, v$ directions, 
	whereas in the radial $r$ directions, Dirichlet boundary conditions are imposed. The problem dependent functions $n_0$ and $T_e$  only depend on $r$.

	\section{Numerical methods \label{sec:method}}
	In this section we will first propose our exponential integrator based numerical method (section \ref{sec:expint}) and then the splitting/semi-Lagrangian based approach (section \ref{sec:splitting}). Finally, we discuss the computational cost of both methods (section \ref{sec:cost}).
	
	\subsection{Exponential integrator based scheme \label{sec:expint}}
	\subsubsection{Semi-discretization in time \label{sec:time-disc}}
	Let us note that the periodic variable $z$ is usually responsible for the most stringent CFL condition. Thus, our goal is to derive a numerical method which eliminates this CFL condition. To that end we write equation (\ref{dk}) as follows
	\[ \partial_t f + v \partial_z f = F(f), \]
	where
	\[ F(f) = \frac{\partial_\theta \phi}{r}\partial_rf-\frac{\partial_r \phi}{r}\partial_\theta f + \partial_z\phi\partial_{v}f. \]
	Next, we perform a Fourier transform in the $z$-direction which gives
	\[ \partial_t \hat{f} + ikv \hat{f} = \mathcal{F}\left(F(f)\right), \]
	where $\hat{g}=\mathcal{F}(g)=\hat{g}(t, r, \theta, k, v)$ denotes the Fourier transform of an arbitrary 
	function $g(t, r, \theta, z, v)$ in $z$. The associated variable in frequency space is denoted by $k$.
	Using the variation of constants formula, the exact solution can then be written as 
	$$
	\hat{f}(t)=e^{-ikvt}\hat{f}(0) +\int_0^t e^{ikv(s-t)} \mathcal{F} \left( {F(f(s))} \right) \,ds, 
	$$
	or equivalently for a time step of size $\Delta t$ from $t^n=n\Delta t$ to $t^{n+1}=(n+1)\Delta t$
	\begin{align}
		\hat{f}(t^{n+1})&=e^{-ikv\Delta t}\hat{f}(t^n) \nonumber\\
		\label{duhamel}
		&\qquad + e^{-ikv\Delta t} \int_0^{\Delta t} e^{ikvs} \mathcal{F} \left( {F(f(t^n+s))} \right) \,ds. 
	\end{align}
	To obtain a first order exponential integrator we approximate $f(t^n+s)$ by $f(t^n)$ and 
	introduce the $\varphi_\ell$-functions defined by the recurrence relation
	$$
	\varphi_\ell(z)=z\varphi_{\ell+1}(z)+\frac{1}{\ell!}
	$$
	with $\varphi_0(z)=e^z$. The first three $\varphi_\ell$-functions are
	$$
	\varphi_1(z)=\frac{e^z-1}{z},\quad\varphi_2(z)=\frac{e^z-1-z}{z^2},\quad\varphi_3(z)=\frac{e^z-1-z-\frac{z^2}{2}}{z^3}.
	$$
	Consequently, we can write the first order exponential Euler method in the following form (denoting by $f^n$ an approximation to $f(t^n)$)
	\[ \hat{f}^{n+1} = e^z\hat{f}^n+\Delta t\varphi_1(z) \mathcal{F} \left( {F(f^n)} \right) \]
	for $z=-ikv\Delta t$.
	Note that for $v \to 0$ 
	we recover the explicit Euler method
	\[ \hat{f}^{n+1} = \hat{f}^{n} + \Delta t \mathcal{F}(F(f^n)) \]
	which makes it clear that, while we have eliminated the CFL condition in $z$, we still have to satisfy the CFL condition corresponding to $F(f^n)$.
	
	A significant literature has been dedicated to exponential integrators (see for example \cite{cox, ei}). In particular, order conditions have been derived that allow for the construction of higher order methods.
	\color{black} In 1987 Strehmel and Weiner \cite{strehmel} introduced an exponential integrator of second order by using  rational functions rather than $\varphi$-functions. To the best of our knowledge, the first systematic approach to exponential integrators using $\varphi$-functions was done in \cite{hochbruck} by Hochbruck \& Ostermann. In the present work we will mostly consider the two-stage second order scheme from  \cite{hochbruck} \color{black}
	\begin{subequations}
		\begin{align}
			\hat{k}_1&=e^{z}\hat{f}^n+\Delta t\varphi_1(z) \mathcal{F} \left( {F(f^n)} \right) \\
			\hat{f}^{n+1}&=e^{z}\hat{f}^n+\Delta t\left[\left(\varphi_1(z)-\varphi_2(z)\right) \mathcal{F} \left( {F(f^n)} \right) +\varphi_2(z) \mathcal{F} \left( {F(k_1)} \right) \right]
		\end{align}
		\label{ei_2}
	\end{subequations}
	and the four-stage fourth order scheme, namely the Cox-Matthews scheme \cite{cox} 
	\begin{eqnarray}
		\hat{k}_1&=&e^{\frac{1}{2}z}\hat{f}^n+\frac{\Delta t}{2}\varphi_1\left(\tfrac{1}{2}z\right) \mathcal{F} \left( F(f^n) \right) \nonumber\\
		\hat{k}_2&=&e^{\frac{1}{2}z}\hat{f}^n + \frac{\Delta t}{2} \varphi_1\left(\tfrac{1}{2}z\right) \mathcal{F} \left( {F(k_1)} \right)
		\nonumber\\
		\hat{k}_3&=&e^{z}\hat{f}^n + \Delta t \left[ \varphi_1\left(\tfrac{1}{2}z\right) \mathcal{F} \left( {F(k_2)} \right) + \frac{1}{2}  \varphi_1 \left( \tfrac{1}{2}z \right) \left( e^{\frac{1}{2}z} -1 \right) \mathcal{F} \left( F(f^n) \right) \right]
		\nonumber\\
		\hat{f}^{n+1}&=&e^{z}\hat{f}^n+\Delta t \Big[\left. \left( \varphi_1(z)-3\varphi_2(z)+4\varphi_3(z) \right) \mathcal{F} \left( {F(f^n)} \right) \right.
		\nonumber\\
		&&+   \left( 2\varphi_2(z)-4\varphi_3(z)\right) \left( \mathcal{F} \left( {F(k_1)} \right) +\mathcal{F} \left( {F(k_2)} \right) \right) 
		\nonumber\\
		&&+ \left. \left( 4\varphi_3(z)-\varphi_2(z) \right) \mathcal{F} \left( {F(k_3)} \right) \Big]\right..
		\label{ei_4}
	\end{eqnarray}

	\subsubsection{Phase space approximation \label{eq:space-discretization}} 
	To complete the numerical scheme, one has to detail the phase space approximation. That is, we have to specify how to discretize the explicit part of the exponential integrator in space. In the proposed algorithms, one has to approximate $F(f(t^n))$ with respect 
	to the $r$, $\theta$, and $v$ directions. Using the notation of the advection formulation (\ref{dk}), the right hand side is of the form
	\begin{align*}
		F(f^n)=\frac{1}{r} \{ \phi ^n, f^n \}+\partial _z\phi^n\partial _vf^n,
	\end{align*}
	where the Poisson bracket is given by $\{ \phi ^n, f^n \}=\partial _rf^n\partial _\theta \phi^n - \partial _\theta f^n\partial_r \phi^n$ and is discretized by the second order Arakawa finite difference scheme \cite{arakawa1966}. The remaining part $\partial _z\phi^n\partial _vf^n$ 
	is discretized by a second order upwind scheme. Consequently, we obtain a CFL type condition for the  $r, \theta,$ and $v$ directions, but not in the stiffest $z$-direction.
	
	Let us discuss the choice of the finite difference scheme to discretize the Poisson bracket in more detail. The first idea could be to use
	\[ \{\phi^n, f^n\} \approx D_r(f^n)D_{\theta}(\phi^n) - D_{\theta}(f^n)D_r(\phi^n), \]
	where $D_r$ and $D_{\theta}$ are the standard centered difference approximations of the first derivative. The resulting scheme is second order accurate and conserves mass exactly. Unfortunately, the present scheme is not as favorable with respect to momentum and energy conservation. In \cite{arakawa1966}, Arakawa introduced a finite difference method that can simultaneously conserve mass, momentum, and energy. This numerical scheme considers an equal superposition of three parts
	\begin{subequations}
		\begin{align*}
			\{\phi^n, f^n\} &\approx \tfrac{1}{3}(J^{++} + J^{+x} + J^{x+}) \\
			J^{++} &= D_{\theta}(\phi^n) D_r(f^n) - D_r(\phi^n) D_{\theta}(f^n) \\
			J^{+x} &= D_{\theta}(\phi^n D_r(f^n)) - D_r(\phi^n D_{\theta}(f^n)) \\
			J^{x+} &= D_r(D_{\theta}(\phi^n) f^n) - D_{\theta}(D_r(\phi^n) f^n).
		\end{align*}
	\end{subequations}
	Each of the stencils $J^{++}$, $J^{+x}$, and $J^{x+}$ are mass conservative on their own 
	but only their sum conserves momentum and energy as well (in the case of periodic boundary conditions).
	Thus, in our case this method 
	is attractive because the only error made in these quantities is due to the time discretization 
	(i.e. due to the exponential integrator employed). Since it is known that exponential integrators
	do conserve mass (see the proof in \cite{einkemmer2015KP}) the same is true for the numerical method proposed here. 
	Let us also note that this scheme 
	has been generalized to discontinuous Galerkin type methods and thus higher order variants 
	can be easily constructed \cite{arakawadg2014}. However, since we will focus on time 
	integration in this paper we will not consider such methods further.
	
	Although this is not commonly acknowledged in the literature, even for homogeneous Dirichlet boundary conditions mass is \textit{not} conserved up to machine precision.
	Let us investigate this in more detail. We will proceed as follows. First we show that $J^{++}$ conserves mass to machine precision. This is a straightforward but rather tedious calculation. Then we will show why mass conservation is violated for $J^{+x}$. This result will also allow us to suggest a procedure to fix this deficiency. To do so, we introduce a mesh in $r$ and $\theta$:  
	$r_i=r_{\min} + i h_r$ for $i=0, \dots, N+1$ ($h_r = (r_{\max}-r_{\min})/(N+1)$) 
	and $\theta_j=jh_\theta$ for $j=0, \dots, N$ ($h_\theta=2\pi/N$); moreover, 
	we define $\phi_{i,j}$ (resp. $f_{i,j}$) as an approximation of $\phi(r_i, \theta_j)$ (resp. of $f(r_i, \theta_j)$).

	The stencil $J^{++}$ discretizes the Poisson bracket in advection form. We start with the sum over the first term
	$$ \sum_{i=1}^N \sum_{j=1}^N [D_{\theta}(\phi) D_r(f)]_{i,j} = \frac{1}{4 h_r h_{\theta}} \sum_{i=1}^N \sum_{j=1}^N (\phi_{i,j+1}-\phi_{i,j-1})(f_{i+1,j}-f_{i-1,j}), $$
	where, for simplicity, we have assumed that both the $r$ and the $\theta$-directions are discretized using $N$ grid points. %
	In addition, we have dropped the superscript $n$, which denotes the value at the previous time step, from both $\phi$ and $f$. 
	Let us further note that we do not consider the volume element (arising from cylindrical coordinates) in the calculation. This is valid since the Poisson bracket is multiplied by $1/r$ which immediately annihilates the volume element. Therefore, all results obtained here apply equally to cylindrical and cartesian coordinates.
	
	The homogeneous Dirichlet boundary condition in the $r$-direction is then enforced by setting $f_{0,j}$, $f_{N+1,j}$, $\phi_{0,j}$, and $\phi_{N+1,j}$ (the ghost cells) to zero for all $j$. Using these relations and applying summation by parts in the $r$-direction we obtain
	$$ 4h_r h_{\theta} \sum_{i=1}^N \sum_{j=1}^N [D_{\theta}(\phi) D_r(f)]_{i,j} = 
	\sum_{j=1}^{N}\left[\sum_{i=1}^{N}f_{i,j}(\phi_{i-1,j+1}-\phi_{i-1,j-1})-\sum_{i=1}^{N}f_{i,j}(\phi_{i+1,j+1}-\phi_{i+1,j-1})\right]. $$
	Now, integration by parts in the $\theta$-direction is straightforward as periodic boundary conditions are imposed in that direction. We obtain
	$$ \sum_{i=1}^{N}\sum_{j=1}^{N}(\phi_{i+1,j}-\phi_{i+1,j})(f_{i,j+1}-f_{i,j-1}) 
	= 4h_r h_{\theta} \sum_{i=1}^{N}\sum_{j=1}^{N} [D_r(\phi) D_{\theta}(f)]_{i,j}. $$
	Since this is exactly the sum over the second term in $J^{++}$, we conclude at once that mass is conserved up to machine precision.
	Now, let us consider $J^{+x}$. Since we have periodic boundary conditions in the $\theta$-direction we get
	\[ \sum_{i=1}^N \sum_{j=1}^N [D_{\theta}(\phi D_r(f))]_{i,j} = 0. \]
	For the second term we have (focusing only on the summation in $i$) 
	\[ \sum_{i=1}^N [D_r(\phi D_{\theta}(f))]_{i,j}
	= \frac{1}{2h_r} \sum_{i=1}^N (g_{i+1,j}-g_{i-1,j})
	= \frac{1}{2h_r}( g_{N+1,j} + g_{N,j} - g_{1,j} - g_{0,j}),
	\]
	where $g = \phi D_{\theta}(f)$. The only local choice to ensure conservation of mass is
	\[ g_{0,j} = -g_{1,j}, \qquad g_{N+1,j} = -g_{N,j}, \quad \forall j=1, \dots, N. \]
	This should not be that surprising. In fact, showing mass conservation for the standard second order finite volume scheme proceeds very similar to the previous argument. In this case the condition above has the natural interpretation that $g$ is zero at the cell interface (in our notation $g_{1/2,j} = \frac{1}{2}(g_{0,j} + g_{1,j}) = 0, \, \forall j=1, \dots, N$). The above calculation gives the same result if applied to $J^{x+}$. %
	
	Note that strictly speaking we have assumed here that $J^{+x}$ and $J^{x+}$ conserve mass separately. This is, of course, not necessarily true as the error in mass committed by $J^{+x}$ could be compensated for by $J^{x+}$. To check this possibility we consider (there is no need to take $J^{++}$ into account as that term conserves mass exactly)
	\[ \sum_{i=1}^N \sum_{j=1}^N (J^{+x}_{i,j} + J^{x+}_{i,j})
	= \sum_{i=1}^N \sum_{j=1}^N ([D_r(f D_{\theta}(\phi))]_{i,j} - [D_r(\phi D_{\theta}(f))]_{i,j}). \]
	Enforcing the Dirichlet boundary conditions and performing summation by parts gives (note that, as before, we define $g = \phi D_{\theta}(f)$)
	\[ \sum_{i=1}^N \sum_{j=1}^N (J^{+x}_{i,j} + J^{x+}_{i,j})
	= \frac{1}{h_r} \sum_{j=1}^N (g_{1,j} - g_{N,j}) \]
	which is not zero in general.

	The previous discussion leaves us with the realization that no 
	matter how we enforce the boundary conditions in at least one of the stencils, mass 
	conservation will be violated at the boundary.
	We can, however, remedy this deficiency by enforcing different boundary conditions for the different stencils. This will introduce a first order error at the boundary but will ensure mass conservation up to machine precision. We will investigate this issue further in section \ref{sec:machineprec}, where we will find that this approach also has a positive effect on energy conservation but a negative effect on $L^2$ norm conservation.

	\subsection{Splitting/semi-Lagrangian method \label{sec:splitting}}
	\label{BSL}
	In view of the numerical comparison conducted in the next section, we recall the basic steps of the algorithm used to approximate the drift-kinetic 
	model (\ref{dk}) coupled with the quasi-neutrality equation \eqref{qn} by means of the splitting/semi-Lagrangian method (see \cite{vlasovia, grandgirard}). 

	\medskip
	
	\noindent Assuming $f^n$ and $\phi^n$ some known approximations to $f(t^n)$ and $\phi(t^n)$ (with $t^n=n\Delta t$), 
	the main steps of the algorithm to compute $f^{n+1}$ and $\phi^{n+1}$ are 
	\begin{itemize}
		\item solve the $1D$ advection $\partial_t f+v\partial_z f=0$ with step size $\Delta t/2$ 
		using a semi-Lagrangian method,  %
		\item solve the $1D$ advection $\partial_t f-\partial_z\phi^n \partial_v f=0$ with step size $\Delta t/2$ 
		using a semi-Lagrangian method to get $f^\star$,%
		\item solve the quasi-neutrality equation (\ref{qn}) with right hand side $\int_{\mathbb{R}} rf^\star dv$ to compute the potential $\phi^\star$, with FFT in $\theta, z$ 
		and second order finite difference in $r$,  
		\item compute the derivatives $(\partial_r \phi^\star, \partial_\theta \phi^\star, \partial_z \phi^\star)$ of $\phi^\star$ 
		with cubic splines, 
		\item solve the $2D$ advection in $(r,\theta)$ with step size $\Delta t$ using a semi-Lagrangian method, %
		\item solve the $1D$ advection $\partial_t f-\partial_z\phi^\star\partial_v f=0$ with step size $\Delta t/2$ 
		using a semi-Lagrangian method, %
		\item solve the $1D$ advection $\partial_t f+v\partial_z f=0$ with step size $\Delta t/2$ 
		using a semi-Lagrangian method, to get $f^{n+1}$ %
		\item solve the quasi-neutrality equation (\ref{qn})  with right hand side $\int_{\mathbb{R}} rf^{n+1} dv$, to compute the potential $\phi^{n+1}$, with FFT in $\theta, z$ 
		and second order finite difference in $r$.   
	\end{itemize}
	Contrary to our method, this approach is not restricted by any CFL condition. 
	However, as we shall see, our method has some advantages from a computational point of view. Moreover, 
	for this drift-kinetic model, the semi-Lagrangian method is not conservative (due to the $2D$ spline interpolation in the $(r, \theta)$ plane used in \cite{vlasovia, grandgirard}). Thus the corresponding long time behavior of the numerical solution may become unsuitable. %
	
	\subsection{Computational cost \label{sec:cost}}
	
	For the actual implementation it is advantageous to rewrite the second order exponential integrator given in (\ref{ei_2}) in the following form 
	\begin{align*}
		\hat{k}_1&=  \hat{f}^n+\Delta t \varphi_1(z) \left( \mathcal{F}\left({F(f^n)}\right) - ikv\hat{f}^n \right)\\
		\hat{f}^{n+1}&=\hat{k}_1+\Delta t\varphi_2(z) \Big[ \mathcal{F}\Big({F(k_1)} -{F(f^n)}\Big) \Big]. 
	\end{align*}
	In this format we see that the method requires four FFTs (compute $f^n$ and $k_1$ from $\hat{f}^n$ and 
	$\hat{k_1}$, compute $\mathcal{F}({F(f^n)})$ and $\mathcal{F}({F(k_1)})$ from $F(f^n)$ and $F(k_1)$).
	
	In addition, we require the evaluation of two right hand sides which consists of a stencil code and the 
	quasi-neutrality equation (\ref{qn}). The latter in turn requires two FFTs and the Thomas algorithm 
	(which scales as $\mathcal{O}(N_r)$, with $N_r$ the number of points in the $r$ direction) %
	but only takes place in a three-dimensional subspace. 
	Consequently the main numerical effort is determined by the four one-dimensional FFTs that have to be applied to the entire four-dimensional phase space.
	
	For comparison, the semi-Lagrangian approach requires four one-dimensional semi-Lagrangian steps in one dimension and one semi-Lagrangian step in two dimensions. The solution of the quasi-neutrality equation incurs the same computational effort in both schemes. In \cite{vlasovia, grandgirard} the semi-Lagrangian steps are done using cubic spline interpolation. For the constant advection case in the $z$-direction two FFTs could be performed instead of the cubic spline interpolation (note, however, that this is not possible for the remaining semi-Lagrangian steps). It is not entirely straight forward to compare the performance of cubic spline interpolation to FFT based techniques. However, since FFT libraries are available that operate close to the theoretical hardware limit with respect to performance, we might expect that constructing and then evaluating the cubic spline interpolation is at least (which would require that both the construction as well as the evaluation step operate close to the theoretical bandwidth limit) as expensive as two FFTs. 
	Under this assumption the remaining cost of the semi-Lagrangian method results from $2$ one-dimensional advections and a single two-dimensional advection, while the remaining cost of the exponential integrator is due to $3$ evaluations of $F$. We note that the evaluation of $F$ is a standard stencil code for which it is well known that performance comparable to the theoretical peak performance of the hardware can be achieved on virtually all architectures. Consequently, the exponential integrator approach introduced in this paper has a clear advantage from a computational point of view. %
	
	In addition, the splitting approach still has to tackle the two-dimensional semi-Lagrangian step which is a further drawback of this method %
	(both with respect to efficiency but also with respect to ease of implementation). %
	
	\color{black}
	\begin{table}
		\begin{center}
			\begin{tabular}{l|rrr}
				& semi-Lagrangian & perturbation & direct \\
				\hline
				coarse ($\Delta t=10$)& 2196 & 482 & 495 \\
				fine ($\Delta t=4$)& 54122 & 13259 & 13116 \\
			\end{tabular}
		\end{center}
		\caption{\color{black}The run times in seconds for the semi-Lagrangian method (using an Euler solver to determine the feet of the characteristics for the two-dimensional interpolation) and the second order exponential integrator (for the perturbation and the direct formulation) are shown. The coarse problem uses $32 \times 32 \times 32 \times 64$ grid points and integrates up to a final time of $T=8000$. The fine problem uses $64 \times 64 \times 64 \times 128$ grid points and integrates up to a final time of $T=6000$.
			\label{tab:comp}}
	\end{table}
	
    In Table \ref{tab:comp} we directly compare the runtime of the semi-Lagrangian approach to the second order exponential integrator proposed in this paper (the differences between the perturbation formulation and the direct formulation as well as further details about our numerical method are discussed in the next section). Even though we use the same computer (dual socket Intel Intel Xeon E5-2630 v3 with 32GB of main memory) and a sequential run for all the simulations, these numbers have to be considered with care, since the two codes are written in different languages (the semi-Lagrangian code uses the library SeLaLib and is written in Fortran while the integrator proposed here is implemented in C). However, we observe that the proposed method is more than four times faster compared to the semi-Lagrangian approach.
	
	\color{black}
	
	\section{Numerical simulations \label{sec:numerical}}
	In this section, we detail the physical parameters of the considered test case. 
	First, the initial value is given by 
	\begin{align*}
	f(t=0,r,\theta,z,v)&=\\
	&&\hspace{-2.5cm} f_{\rm eq}(r,v)\left[1+\epsilon \exp\left(-\frac{(r-r_p)^2}{\delta r}\right)\cos\left(\frac{2\pi n}{L}z+m\theta\right)\right],
	\end{align*}
	where the equilibrium function is 
	\begin{align}
	\label{feq}
	f_{\rm eq}(r,v)=\frac{n_0(r)\exp(-\frac{v^2}{2T_i(r)})}{(2\pi T_i(r))^{1/2}}.
	\end{align}
	The radial profiles $\{T_i,T_e,n_0\}$ have the analytic expressions
	$$
	\mathcal{P}(r) = C_\mathcal{P}\exp\left(-\kappa_\mathcal{P}\delta r_{\mathcal{P}}\tanh\left(\frac{r-r_p}{\delta r_{\mathcal{P}}}\right)\right), \; \mathcal{P}\in \{T_i,T_e,n_0\}, 
	$$
	where the constants are 
	$$
	\ C_{T_i}=C_{T_e}=1,\ C_{n_0}=\frac{r_{\rm max}-r_{\rm min}}{
		\int_{r_{\rm min}}^{r_{\rm max}}\exp(-\kappa_{n_0}\delta r_{n_0}\tanh(\frac{r-r_p}{\delta r_{n_0}}))\,dr}.
	$$
	Finally, we consider the parameters of \cite{BC2013} (MEDIUM case)
	\begin{align*}
	r_{\min}	&=	0.1,\ r_{{\rm max}}=14.5, \\
	\kappa_{n_{0}}	&=	0.055,\ \kappa_{T_{i}}=\kappa_{T_{e}}=0.27586, \\
	\delta r_{T_{i}}	&=	\delta r_{T_{e}}=\frac{\delta r_{n_{0}}}{2}=1.45,\ \epsilon=10^{-6},\ n=1,\ m=5, \\
	L	&=	1506.759067,\ r_{p}=\frac{r_{{\rm min}}+r_{{\rm max}}}{2},\delta r=\frac{4\delta r_{n_{0}}}{\delta r_{T_{i}}},
	\end{align*}
	and use a $v$-range of 
	$$
	v \in [-7.32,7.32].
	$$
	For the direct formulation, we use the boundary conditions
	$$
	f(r_{\rm min},\theta ,z,v)=f_{eq}(r_{\rm min},v) \qquad
	f(r_{\rm max},\theta ,z,v)=f_{eq}(r_{\rm max},v).
	$$
	Note that these are not homogeneous Dirichlet boundary conditions. It is well known %
	(and supported by our numerical experiments) %
	that the Arakawa scheme works better for homogeneous boundary conditions. In addition to the direct formulation, we therefore also introduce a so-called perturbation formulation.
	First, we note that the equilibrium function $f_{\rm eq}$ defined in (\ref{feq}) is a steady state for our problem. We therefore divide $f$ into
	$$
	f(t,r,\theta ,v)=f_{eq}(r,v)+\delta f(t,r,\theta ,v).
	$$
	With this formulation, our problem (\ref{dk}) becomes
	$$
	\partial_t\delta f+\frac{E_\theta}{r}\partial_r (f_{eq} + \delta f)-\frac{E_r}{r}\partial_\theta \delta f+v\partial_z\delta f+E_z \partial_v (f_{eq} + \delta f)=0,
	$$
	where $E_\theta=-\partial_\theta \phi$, $E_r=-\partial_r \phi$ and $E_z=-\partial_z \phi$. Expanding the various terms we obtain
	$$
	\partial_t\delta f+\frac{E_\theta}{r}\partial_r\delta f-\frac{E_r}{r}\partial_\theta \delta f+v\partial_z\delta f+ E_z \partial_v \delta f
	+\frac{E_\theta}{r}\partial_r f_{eq} + E_z \partial _v f_{eq}=0
	$$
	which can be written as
	$$
	\partial_t\delta f + v\partial_z \delta f -F(\delta f) +\frac{E_\theta}{r}\partial_r f_{eq} +E_z\partial _v f_{eq} = 0.
	$$
	Note that the equation is very similar to equation (\ref{dk}). We, however, have obtained two additional source terms, which depend on the equilibrium distribution $f_{eq}$ as well as on the electric field.
	Furthermore, the right hand side of the quasi-neutrality equation (\ref{qn}) becomes
	$$
	\frac{1}{n_0}\int f_{eq} \,dv+\frac{1}{n_0}\int \delta f \,dv-1 = \frac{1}{n_0}\int \delta f \,dv.
	$$
	Due to the similarity of the direct formulation and the perturbation formulation, the same code can 
	be used for both by simply exchanging the right hand side of the quasi-neutrality equation, changing the boundary conditions, and adding the appropriate source terms. Thus, to implement the exponential integrator we consider the following equation 
	$$ \partial_t\delta f + v\partial_z \delta f  = F_{pert}(\delta f),  $$
	with
	$$ F_{pert} = F(\delta f) -\frac{E_\theta}{r}\partial_r f_{eq} - E_z\partial _v f_{eq},  $$
	and proceed as in section \ref{sec:time-disc}  (with $F$ replaced by $F_{pert}$). The space discretization of the source terms can be done either analytically or using a numerical approximation. In our implementation we have used standard centered differences. The Arakawa scheme that is used to discretize $F(\delta f)$ now employs homogeneous Dirichlet boundary conditions for $\delta f$ in the $r$-direction.

	%
	%
	

	\subsection{Numerical results}
	
	In this section, we compare the standard BSL method introduced in section \ref{BSL} to the newly proposed method. 
	To do that, we consider the $4D$ drift-kinetic model in \color{black}cylindrical \color{black}geometry (\ref{dk})-(\ref{qn}). 
	
	We are interested in the time history of the electric energy 
	\begin{equation}
	\label{modephi}
	\sqrt{\int \phi^2(r_p,\theta,z)\,d\theta dz}, 
	\end{equation}
	where $r_p=(r_{\rm max}+r_{\rm min})/2$, the total mass 
	\begin{equation}
	\label{mass}
	{\cal M}(t) = \int_{r_{\rm min}}^{r_{\rm max}} \int_0^{2\pi} \int_0^L \int_{\mathbb{R}} f(t, r, \theta, z, v) r \,dv dz d\theta dr,  
	\end{equation}
	the $L^2$ norm
	\begin{equation}
	\label{l2norm}
	{\cal L}(t) = \sqrt{\int_{r_{\rm min}}^{r_{\rm max}} \int_0^{2\pi} \int_0^L \int_{\mathbb{R}} f^2(t, r, \theta, z, v) r \,dv dz d\theta dr} ,  
	\end{equation}
	and the total energy 
	\begin{align}
	\label{energy}
	{\cal E}(t) &= \int_{r_{\rm min}}^{r_{\rm max}} \int_0^{2\pi} \int_0^L \int_{\mathbb{R}} \frac{v^2}{2} f(t, r, \theta, z, v) r \,dv dz d\theta dr \nonumber\\
	&+ \int_{r_{\rm min}}^{r_{\rm max}} \int_0^{2\pi} \int_0^L \int_{\mathbb{R}} f(t, r, \theta, z, v) \phi(t, r, \theta, z) r \,dv dz d\theta dr.  
	\end{align}
	The quantity in (\ref{modephi}) is known to exponentially increase 
	with time (see, for example, \cite{BC2013, grandgirard}). \color{black}
	The growth rate can be computed by solving the linearized drift-kinetic model (\ref{dk})-(\ref{qn}). This is done by performing a Fourier transform in $(z,\theta)$ and a Laplace transform in the time variable $t$. The resulting dispersion relation is derived in \cite{BC2013}. We have solved this relation using a computer algebra system and obtained the root with the largest imaginary part. This is precisely the (analytically derived) growth rate.\color{black}

	The total mass, $L^2$ norm, and energy are physically conserved quantities in the analytic model. They are therefore an important measure of the physical accuracy of the numerical method.
	
	The following numerical results are shown for two different grid sizes for $(r,\theta ,z,v)$, where the discretization of $32\times 32\times 32\times 64$ is referred to as the coarse grid and the discretization of  $64\times 64\times 64\times 128$ is referred to as the fine grid. For the coarse grid, we integrate up to a final time $T=8000$, while the fine grid is integrated up to $T=6000$. Furthermore, we compare the perturbation formulation with the direct one. 
	
	\subsubsection{Exponential integrator of order 2}
\begin{figure}
		\begin{center}
			\includegraphics[width=0.8\linewidth]{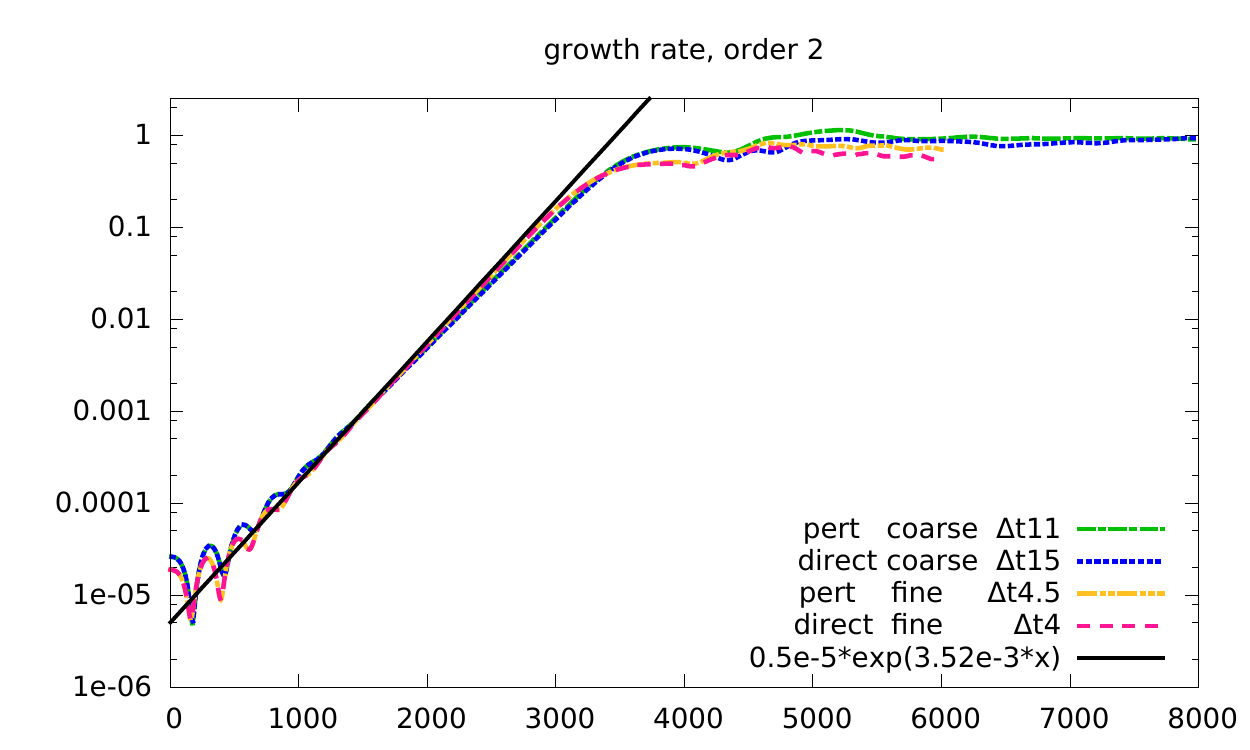}
		\end{center}
		\caption{Time evolution of $\sqrt{\int \Phi^2(r_p,\theta,z)\,d\theta dz}$ with $r_p=(r_{\rm max}+r_{\rm min})/2$ for the exponential integrator of order 2. Compared are the methods obtained by using a fine and a coarse grid as well as the direct formulation against the perturbation formulation.}
		\label{figGR}
	\end{figure}

	In this section, we discuss the numerical results of the second order exponential integrator described by (\ref{ei_2}). 
	In Fig. \ref{figGR}, we plot the time history of (\ref{modephi}) obtained by the perturbation formulation as well as the direct formulation of the exponential integrator for both the coarse and the fine grid. The results are plotted using the maximum possible time step $\Delta t$ for each version. We can see, that all versions reproduce the linear part very well and are close to the analytic growth rate derived in \cite{BC2013}.

	We observe, as expected, that the fine grid requires a smaller time step to maintain stability. However, if we examine the CFL condition a bit more thoroughly we can see that we still can take a relatively large step size compared to the CFL condition in the $z$-direction. The CFL condition is given by
	$$
	\Delta t < \min \left(\text{CFL}_r,\text{CFL}_\theta ,\text{CFL}_z,\text{CFL}_v\right),
	$$
	where the speeds associated with these terms are 
	$$
	v_r= \max_{t,r,\theta,z,v} \left| \frac{E_\theta}{r} \right| ,\quad v_\theta=\max_{t,r,\theta,z,v} \left| \frac{E_r}{r} \right|, 
	\quad v_z=v, \quad v_v= \max_{t,r,\theta,z,v} \left| E_z \right|.
	$$
	Due to the non-linearity of the problem, the CFL condition depends on the numerical solution. It is therefore not straightforward to predict the CFL condition based on theoretical considerations or by numerical data gathered on a coarser grid (see, for example, \cite{filbet2014mixed}). Ideally, an adaptive time stepping scheme would be employed that automatically selects an appropriate step size (such as using the strategy in \cite{gust88} with an embedded exponential integrator \cite{ei}). As we will see in this section this would be extremely beneficial in the linear regime (i.e.~up to approximately $t=3000$) as the nonlinear part of the simulation dictates the time step size. However, since the semi-Lagrangian implementation uses a constant step size we will do the same here. This enables us to perform a more direct comparison with the semi-Lagrangian scheme (in particular, concerning the invariants). However, we consider developing an adaptive version of the present algorithm as future work.

\begin{figure}
		\begin{center}
			\begin{tabular}{cc}
				\includegraphics[width=0.8\linewidth]{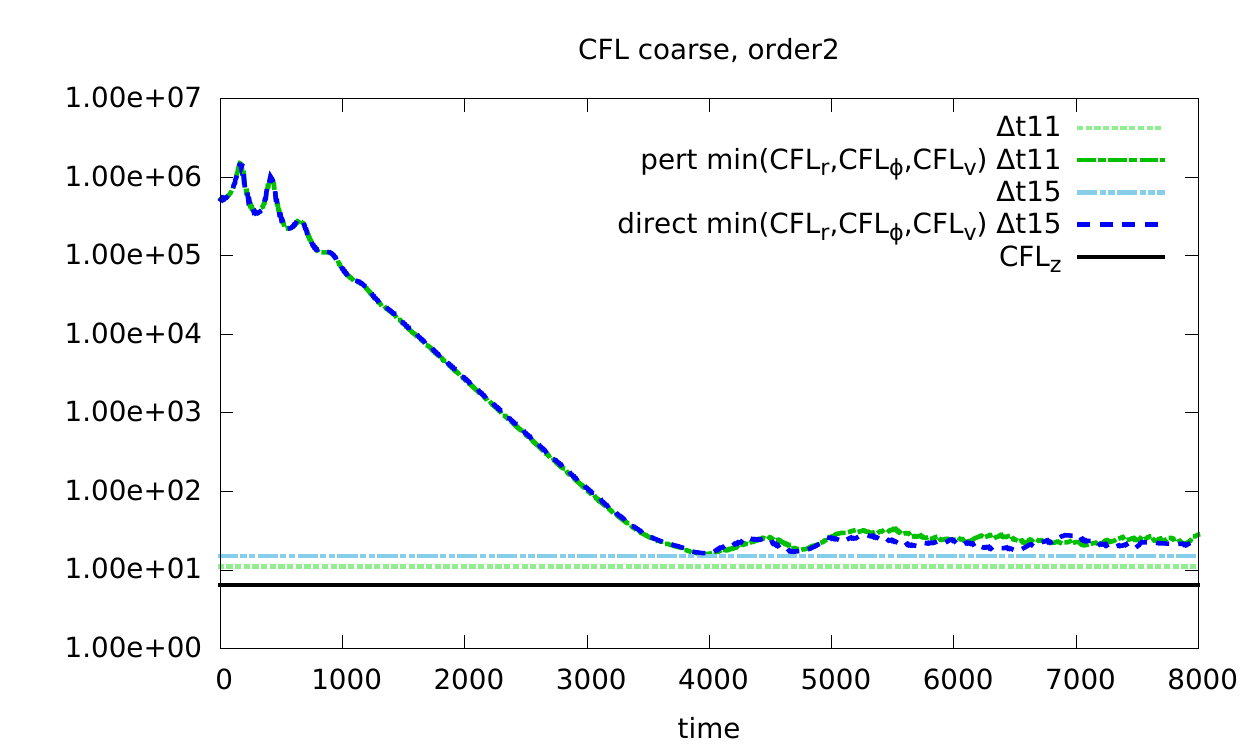} \\
				\includegraphics[width=0.8\linewidth]{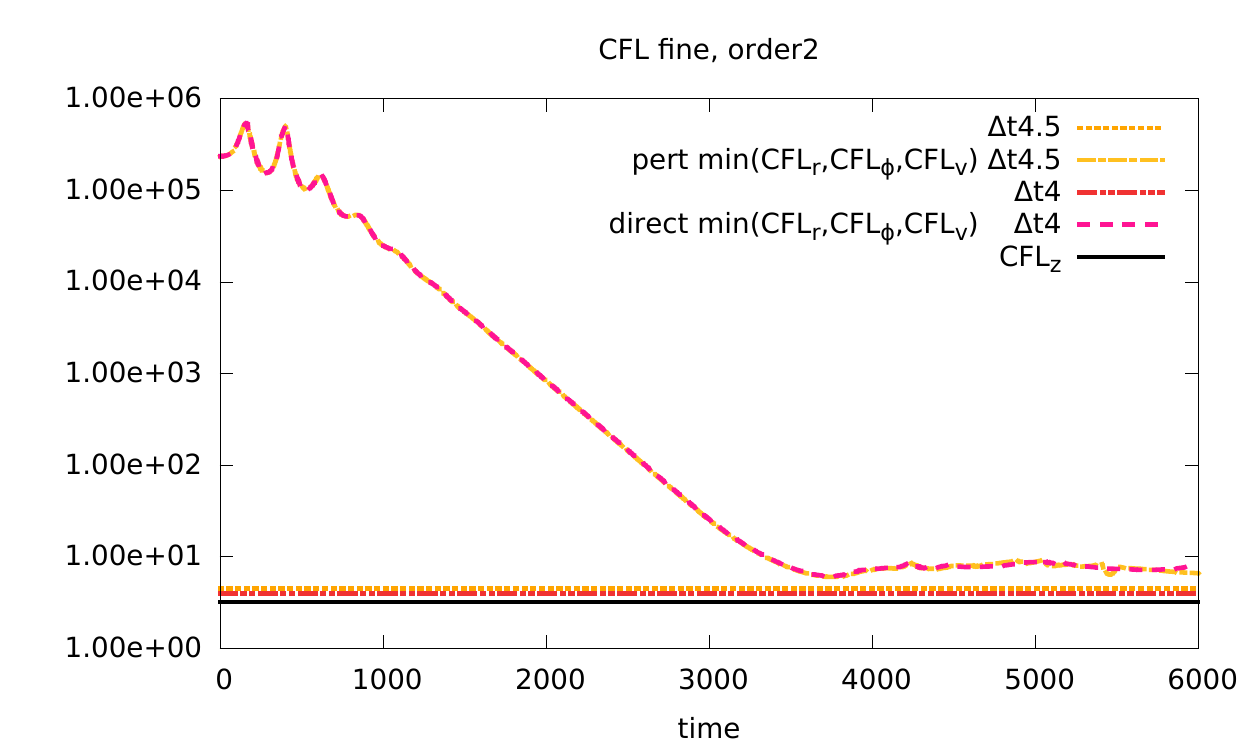}
			\end{tabular}
		\end{center}
		\caption{Comparison between the different CFL numbers as a function of time and the actual time step size. On the top, the coarse problem which uses $32\times 32\times 32\times 64$ grid points is depicted, while on the bottom the fine grid which uses $64\times 64\times 64\times 128$ points is shown.}
		\label{figCFL}
	\end{figure}

	We did evaluate the CFL number during the simulation. 
	For the coarse grid (henceforth denoted with a subscript $c$), the simulation ran until the final time $T=8000$. The computed minimum CFL values for the second order exponential scheme (since these values differ only slightly between the direct and perturbation formulation and change only slightly for reasonable $\Delta t$, we exclusively report the overall minimum) are 
	\begin{align*}
	\text{CFL}_{r,c} = \frac{\Delta r}{v_r} \approx 21 \quad \text{CFL}_{\theta,c} = \frac{\Delta \theta}{v_{\theta}} 
	\approx 15 \quad \text{CFL}_{v,c} = \frac{\Delta v}{v_v} \approx 430000.
	\end{align*}
	Using the maximum value $v=7.32$ and the domain length $L \approx 1506.8$ of our simulation, the CFL condition in the $z$-direction is
	$$
	\text{CFL}_{z,c} = \frac{\Delta z}{v_z} \approx 6.43,
	$$
	which as the smallest value would be the most restrictive CFL condition for the method. However, with the exponential 
	integrator proposed in this paper we are able to take time steps with $\Delta t = 11 > \text{CFL}_{z,c}$ for the perturbation formulation, 
	while still obtaining physically reasonable results. 
	It should be duly noted, however, that we still have to respect the CFL condition in the remaining directions. 
	That is, the exponential integrators have to satisfy $\Delta t < \min (\text{CFL}_{r,c}, \text{CFL}_{\theta,c}, \text{CFL}_{v,c})$. 
	The time history of the CFL number, defined as $\min \left(\text{CFL}_{r, c}, \text{CFL}_{\theta, c}, \text{CFL}_{v, c}\right)$,  
	computed for both the coarse and the fine grid is shown in Fig. \ref{figCFL}. 

	The coarse grid corresponds exactly to the problem considered in \cite{vlasovia}. In that work a time step size of $\Delta t = 8$ is used. That is, we are able to run the simulation with similar (or even larger) time steps while using a significantly cheaper method (see the discussion in section \ref{sec:cost}).
	
\begin{figure}
		\begin{center}
			\begin{tabular}{cc}
				\includegraphics[width=0.8\linewidth]{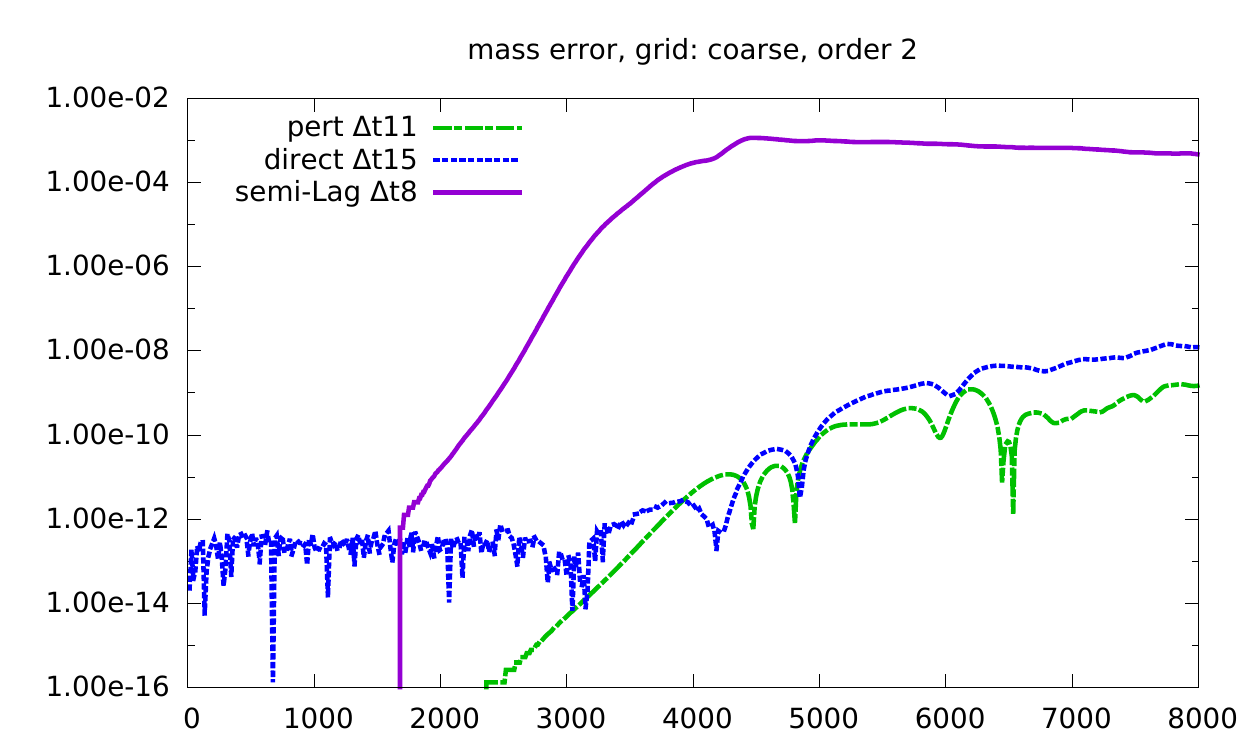} \\
				\includegraphics[width=0.8\linewidth]{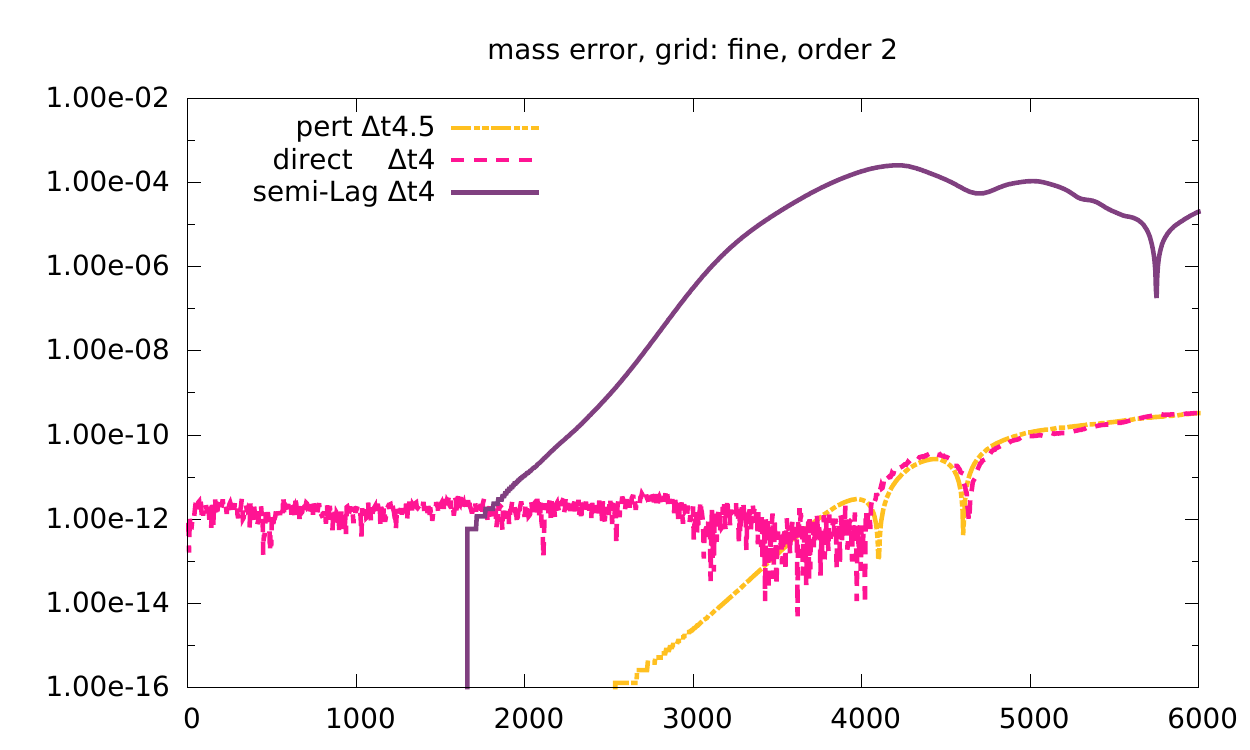}
			\end{tabular}
		\end{center}
		\caption{Comparison of the error in mass for the exponential integrator and the splitting/semi-Lagrangian method. On the top, the coarse problem which uses $32\times 32\times 32\times 64$ grid points is depicted, while on the bottom the fine grid which uses $64\times 64\times 64\times 128$ points is shown. }
		\label{figmass}
	\end{figure}

	For the fine grid, the simulation ran until $T=6000$ and resulted in the values
	\begin{align*}
	\text{CFL}_{r,f} \approx 6 \quad \text{CFL}_{\theta,f}\approx 6 \quad \text{CFL}_{v,f}\approx 180000,
	\end{align*}
	while the computed CFL condition in the $z$-direction is
	$$
	\text{CFL}_{z,f}\approx  3.21, 
	$$
	which again is the most restrictive condition.
	However, in Fig. \ref{figGR}, we see that once again the time step can be chosen larger than that value and our method still produces a stable run. 
	In Fig. \ref{figmass}, we plot the error of the total mass defined in (\ref{mass}) and compare our method with the results from the semi-Lagrangian method. We see that the mass in both the perturbation method as well as the direct formulation is much better preserved than for the semi-Lagrangian approach. 

\begin{figure}
		\begin{center}
			\begin{tabular}{cc}
				\includegraphics[width=0.8\linewidth]{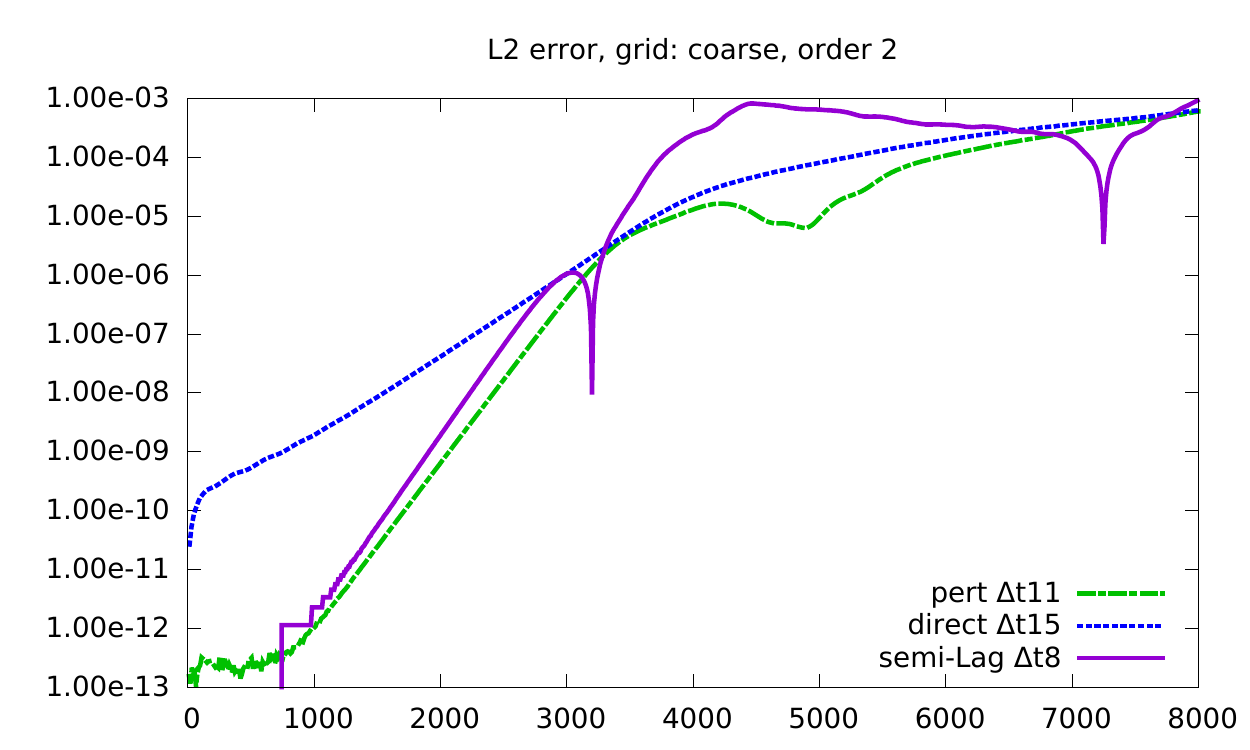} \\
				\includegraphics[width=0.8\linewidth]{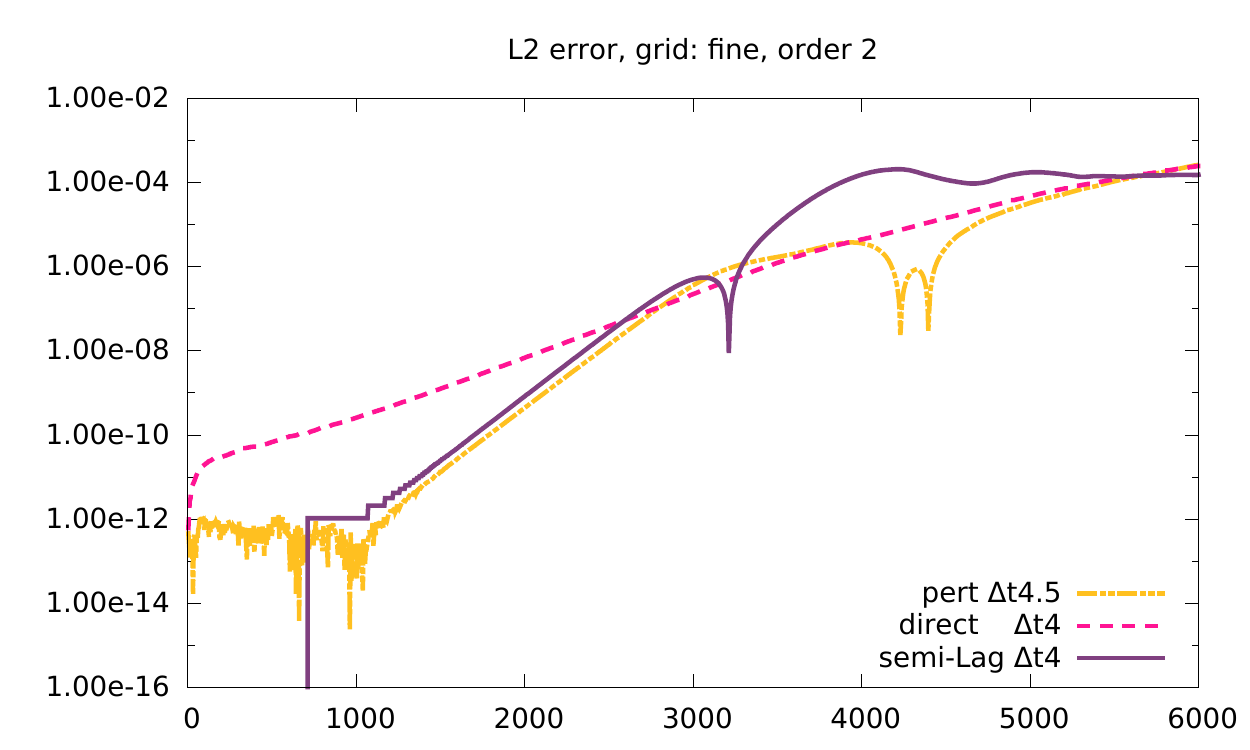}
			\end{tabular}
		\end{center}
		\caption{Comparison of the error in the $L^2$ norm for the exponential integrator and the splitting/semi-Lagrangian method. On the top, the coarse problem which uses $32\times 32\times 32\times 64$ grid points is depicted, while on the bottom the fine grid which uses $64\times 64\times 64\times 128$ points is shown.}
		\label{figL2}
	\end{figure}

	Similarly, the error of the $L^2$ norm defined in (\ref{l2norm}) is plotted in Fig. \ref{figL2}. Here, we can observe, that the error is smaller or equal compared to the semi-Lagrangian method. 
	
\begin{figure}
		\begin{center}
			\begin{tabular}{cc}
				\includegraphics[width=0.8\linewidth]{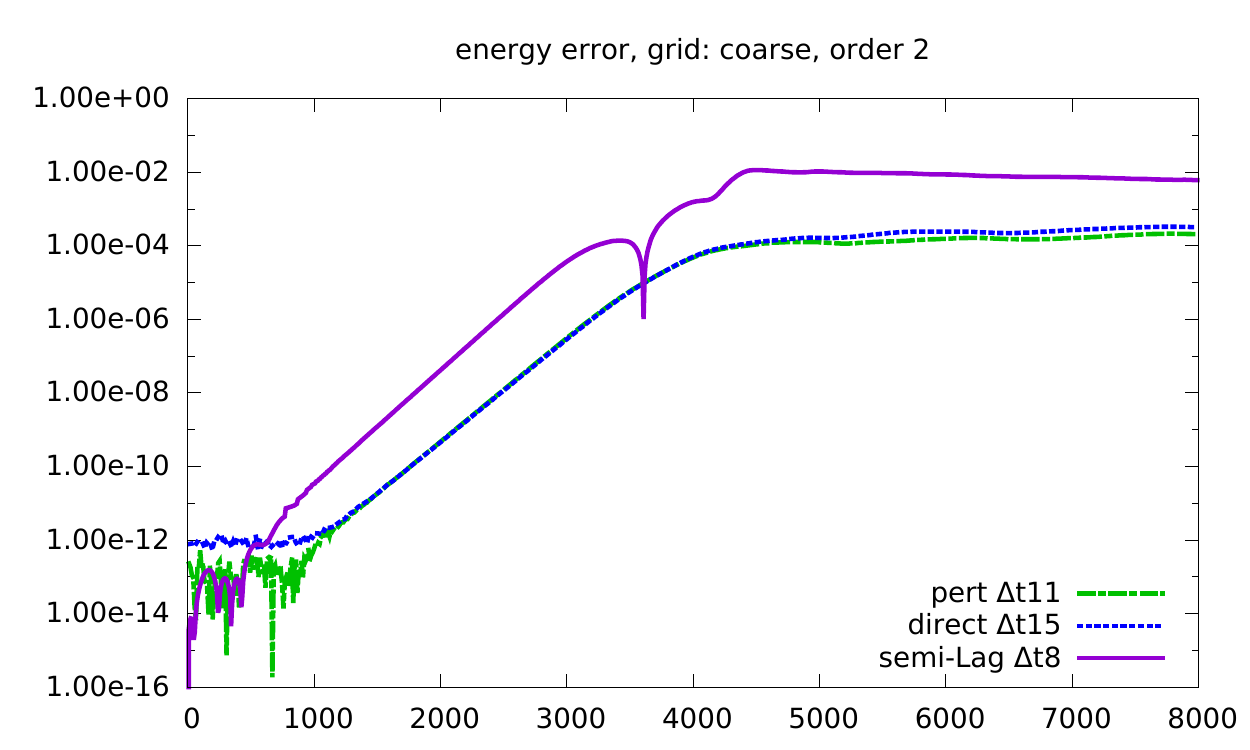} \\
				\includegraphics[width=0.8\linewidth]{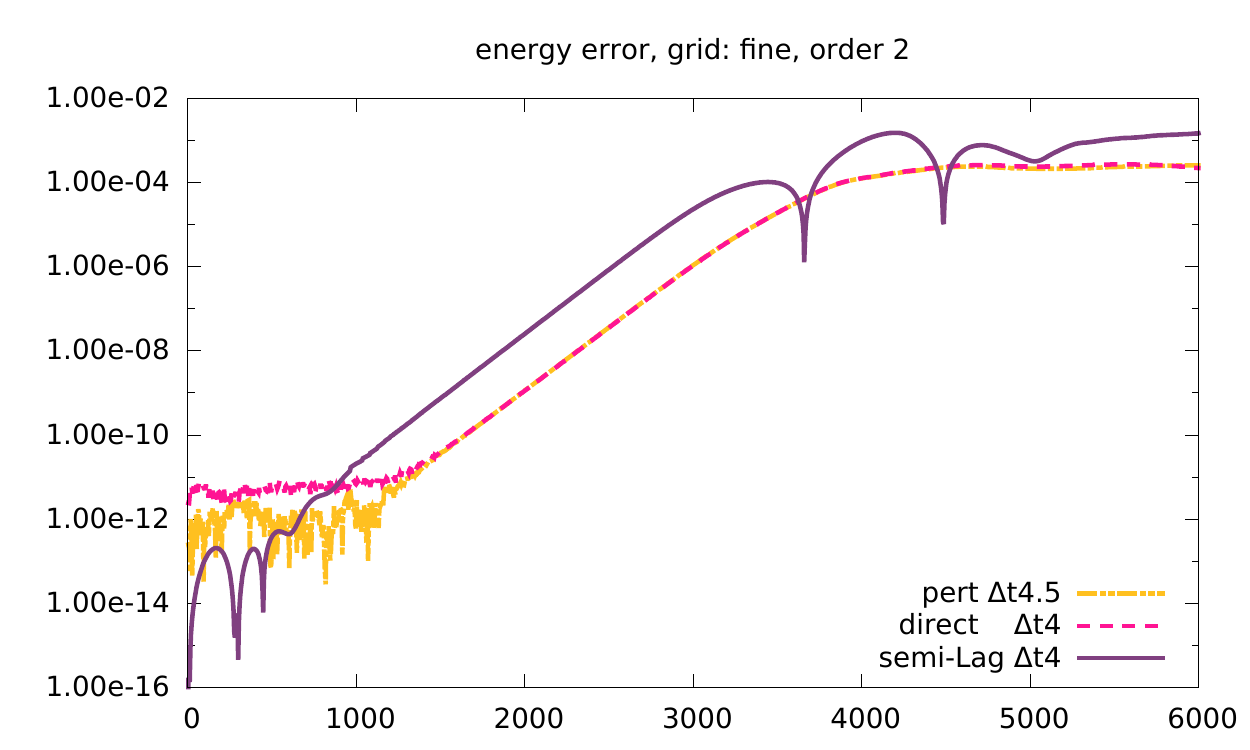}
			\end{tabular}
		\end{center}
		\caption{Comparison of the error in the total energy for the exponential integrator and the splitting/semi-Lagrangian method. On the top, the coarse problem which uses $32\times 32\times 32\times 64$ grid points is depicted, while on the bottom the fine grid which uses $64\times 64\times 64\times 128$ points is shown.}
		\label{figenergy}
	\end{figure}

	In Fig. \ref{figenergy}, the error in the total energy (as defined in (\ref{energy})) is shown. Our method performs approximately an order of magnitude better compared to the semi-Lagrangian scheme. \\

	\subsubsection{Methods of higher order}
	
\begin{figure}
		\begin{center}
			\includegraphics[width=0.8\linewidth]{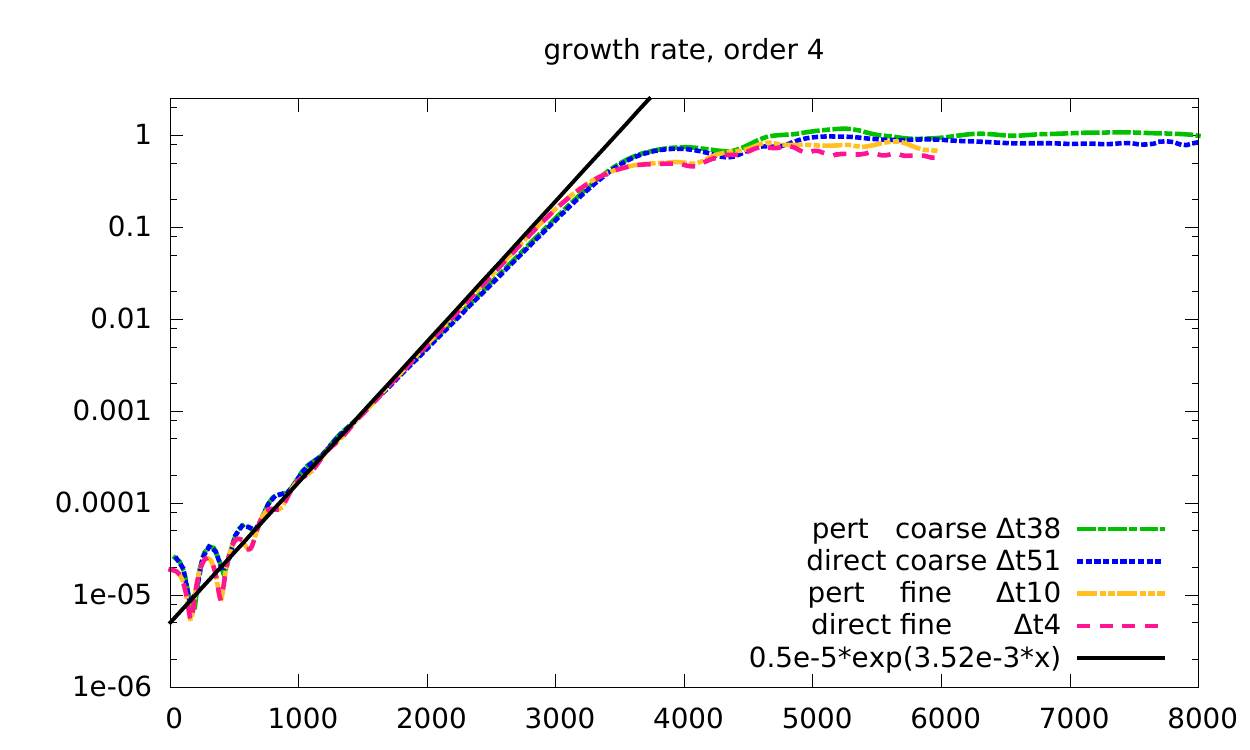}
		\end{center}
		\caption{Time evolution of $\sqrt{\int \Phi^2(r_p,\theta,z) \,d\theta dz}$ with $r_p=(r_{\rm max}+r_{\rm min})/2$ obtained by using a fine and a coarse grid and the exponential integrator of order 4. The direct formulation is compared to the perturbation formulation.}
		\label{figGR_HO}
	\end{figure}

	In this section, we discuss the numerical results for the method of order 4 described in (\ref{ei_4}). In Fig. \ref{figGR_HO}, we can see that all variants of our method, i.e, the perturbation formulation and the direct formulation for both the coarse and the fine grid reproduce the linear growth rate of (\ref{modephi}) (as compared to the analytic result given in \cite{BC2013}). This behavior is similar compared to the second order method.
	
	Using numerical simulations to determine the maximal allowed step size, we observe that the time step size for the fourth order method is significantly less stringent compared to the method of order 2. The maximum allowed time steps for all configurations are listed in Table \ref{tab:stability}. In particular, for the fourth order method in the perturbation formulation we can take significantly larger time steps compared to the second order method. Since all methods are able to resolve the growth phase very well, the ability to take larger time steps is the major advantage of the fourth order method in the linear regime. In addition, employing higher order schemes is useful as they introduce less diffusion and generally provide more accurate results.

	\begin{table}
		\begin{center}
			\begin{tabular}{l|lllll}
				& \multicolumn{2}{c}{coarse} & & \multicolumn{2}{c}{fine} \\
				\cline{2-3} \cline{5-6}
				& direct & perturb. & & direct & perturb. \\
				\hline
				2nd order & 15 & 11 & & 4 & 4.5 \\
				4th order & 51 & 38 & & 4 & 10 \\
			\end{tabular}
		\end{center}
		\caption{The maximum stable time step $\Delta t$ for the second and fourth order scheme is listed for both the coarse and fine grid. \label{tab:stability}}
	\end{table}

\begin{figure}
		\begin{center}
			\begin{tabular}{cc}
				\includegraphics[width=0.8\linewidth]{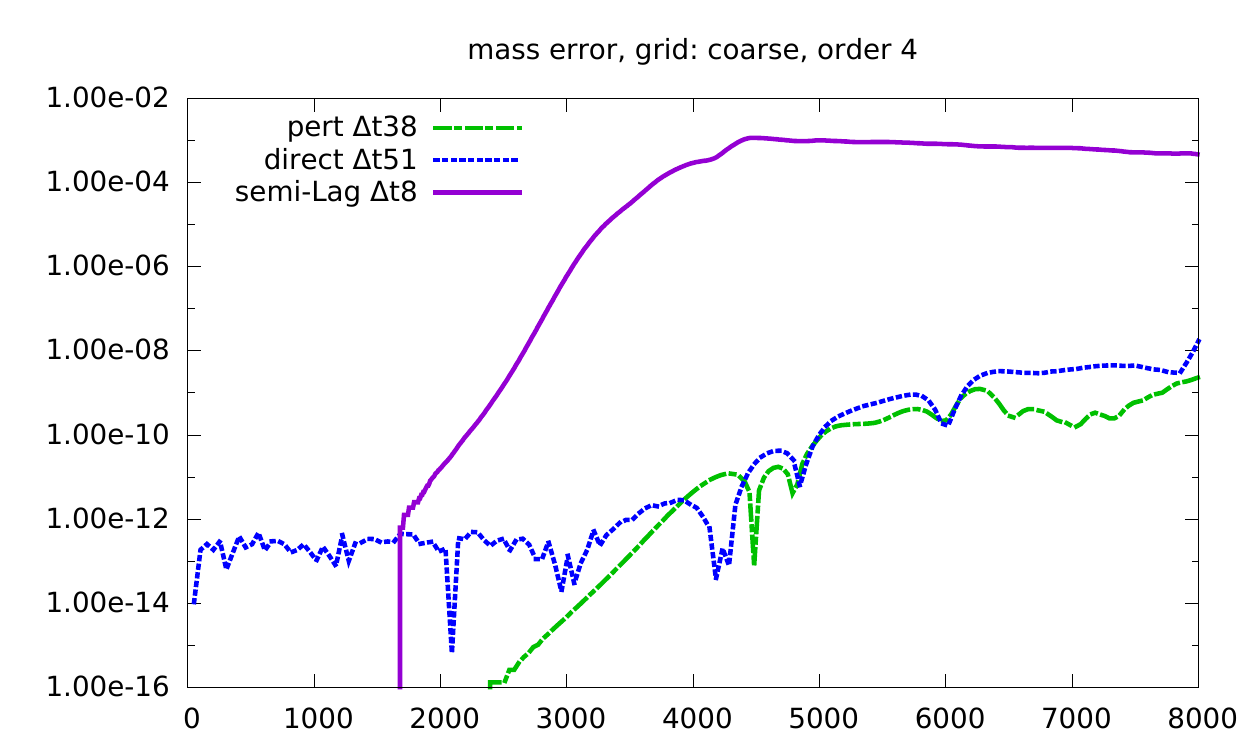} \\
				\includegraphics[width=0.8\linewidth]{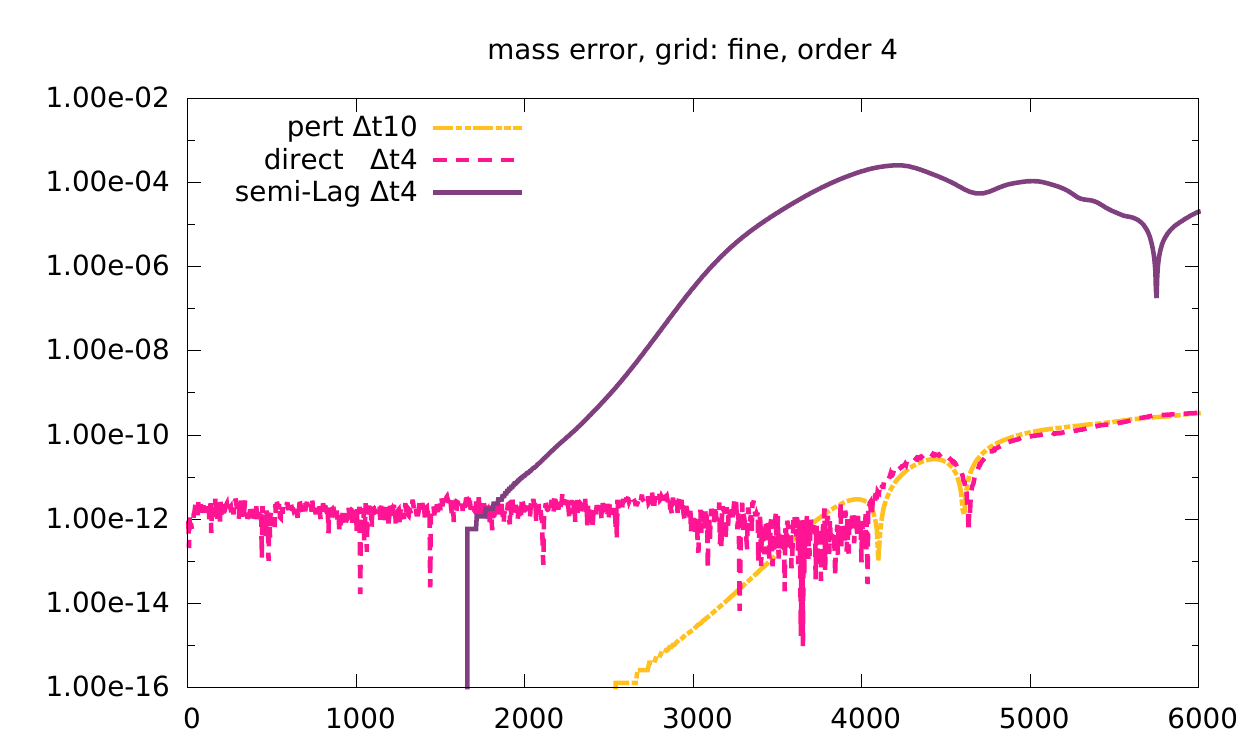}
			\end{tabular}
		\end{center}
		\caption{Comparison of the error in mass for the fourth order exponential integrator and the splitting/semi-Lagrangian method. On the top, the coarse problem which uses $32\times 32\times 32\times 64$ grid points is depicted, while on the bottom the fine grid which uses $64\times 64\times 64\times 128$ points is shown.}
		\label{figmass_o4}
	\end{figure}

	In Fig. \ref{figmass_o4}, we show the error in mass for the fourth order method. The behavior is similar to that of the second order method. In particular, we see a multiple orders of magnitude improvement compared to the semi-Lagrangian approach. This is true even for the significantly larger time steps we are now able to take.

\begin{figure}
		\begin{center}
			\begin{tabular}{cc}
				\includegraphics[width=0.8\linewidth]{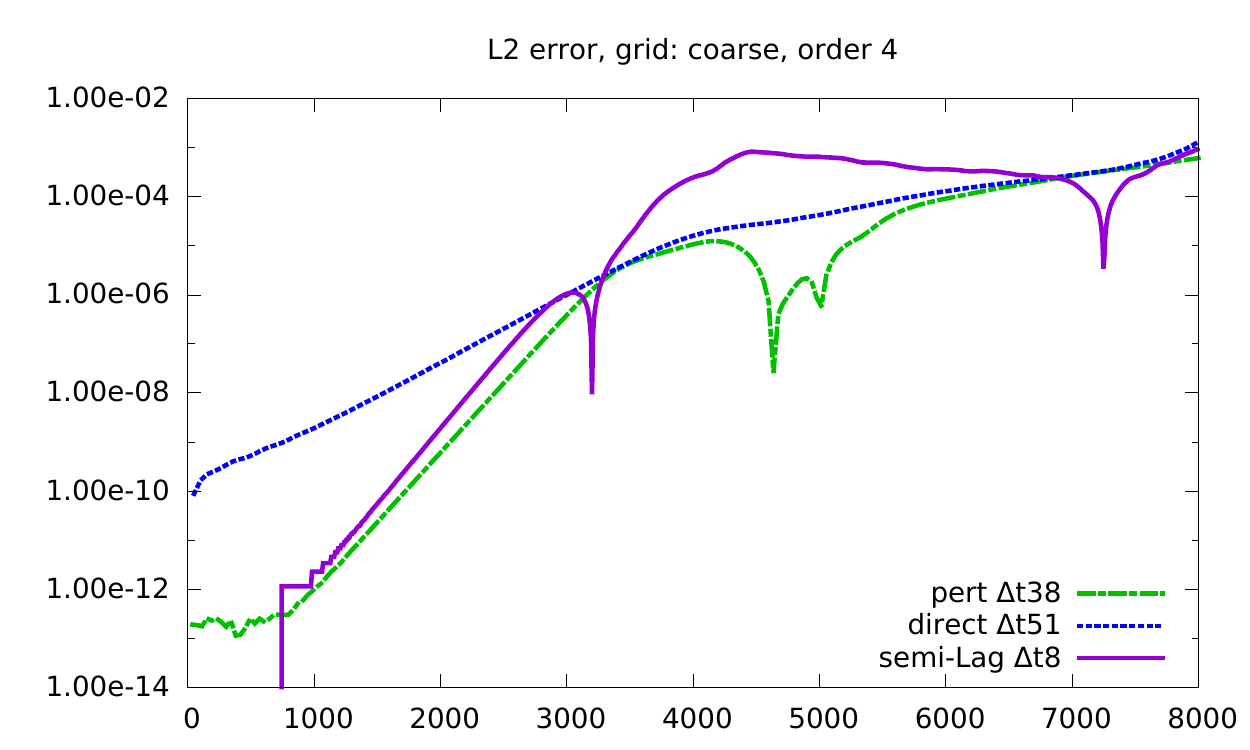} \\
				\includegraphics[width=0.8\linewidth]{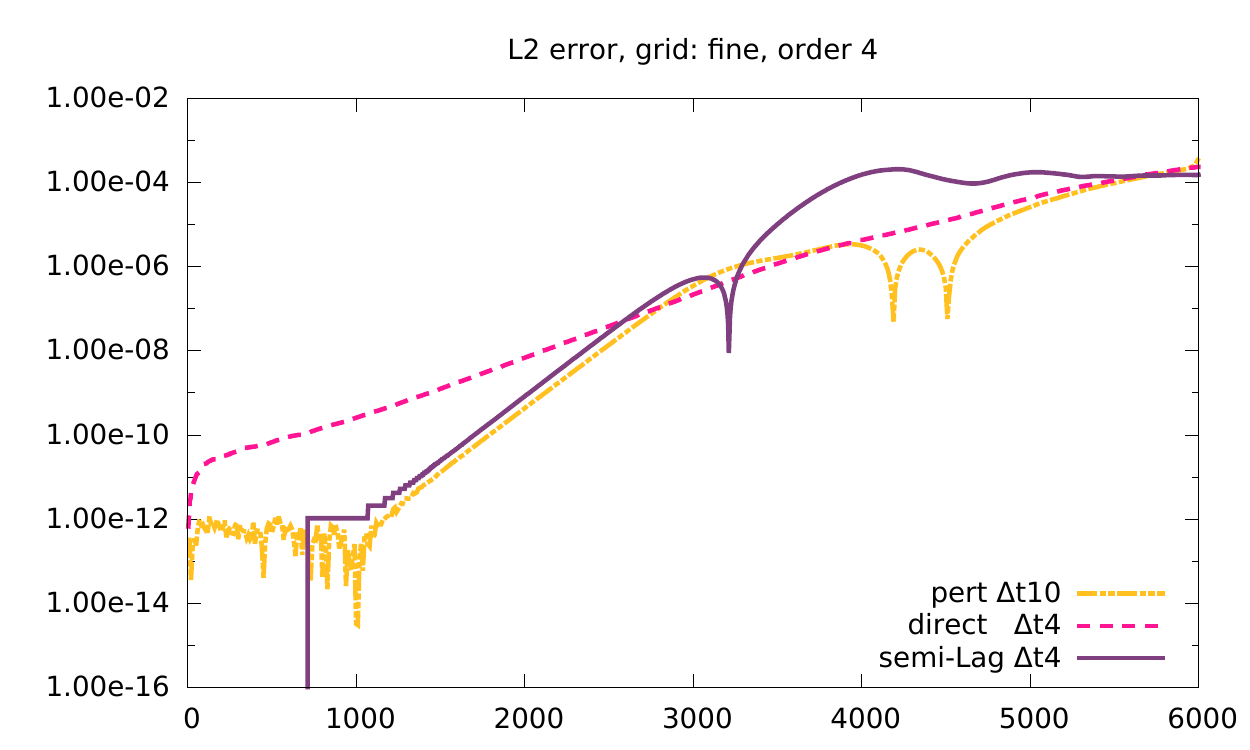}
			\end{tabular}
		\end{center}
		\caption{Comparison of the error in the $L^2$ norm for the fourth order exponential integrator and the splitting/semi-Lagrangian method. On the top, the coarse problem which uses $32\times 32\times 32\times 64$ grid points is depicted, while on the bottom the fine grid which uses $64\times 64\times 64\times 128$ points is shown.}
		\label{figL2_o4}
	\end{figure}

\begin{figure}
		\begin{center}
			\begin{tabular}{cc}
				\includegraphics[width=0.8\linewidth]{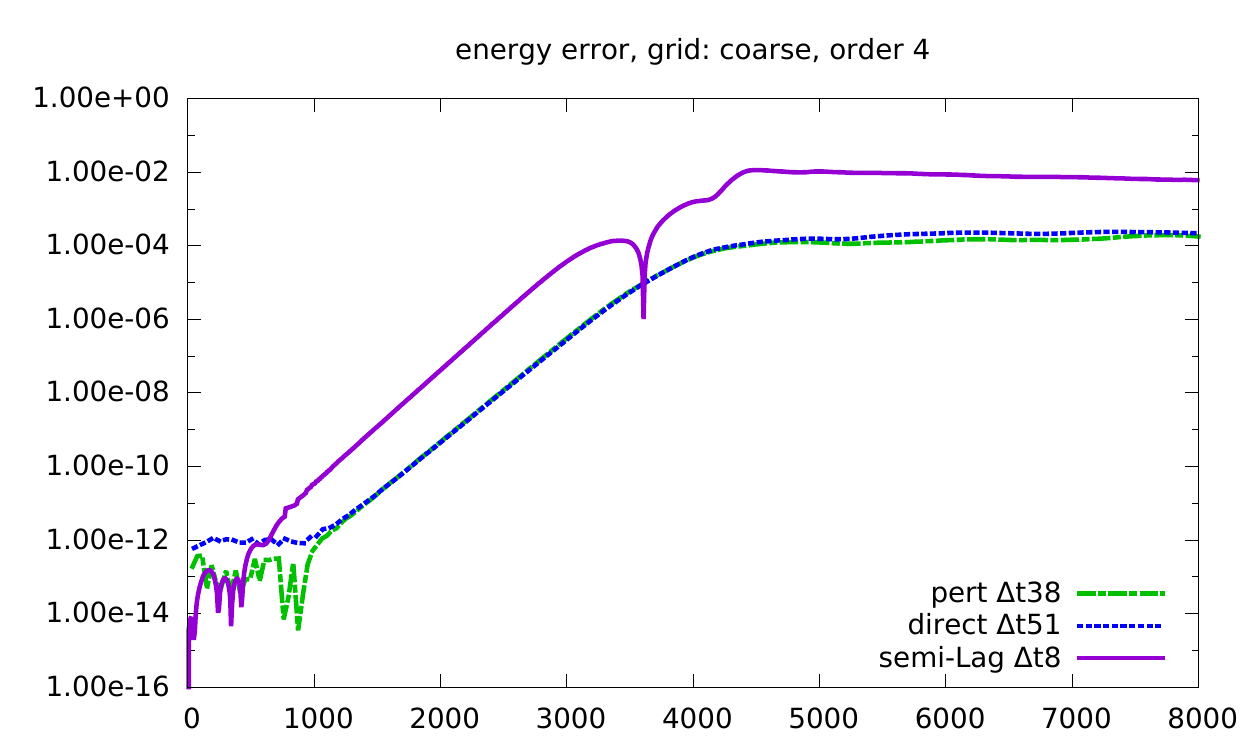} \\
				\includegraphics[width=0.8\linewidth]{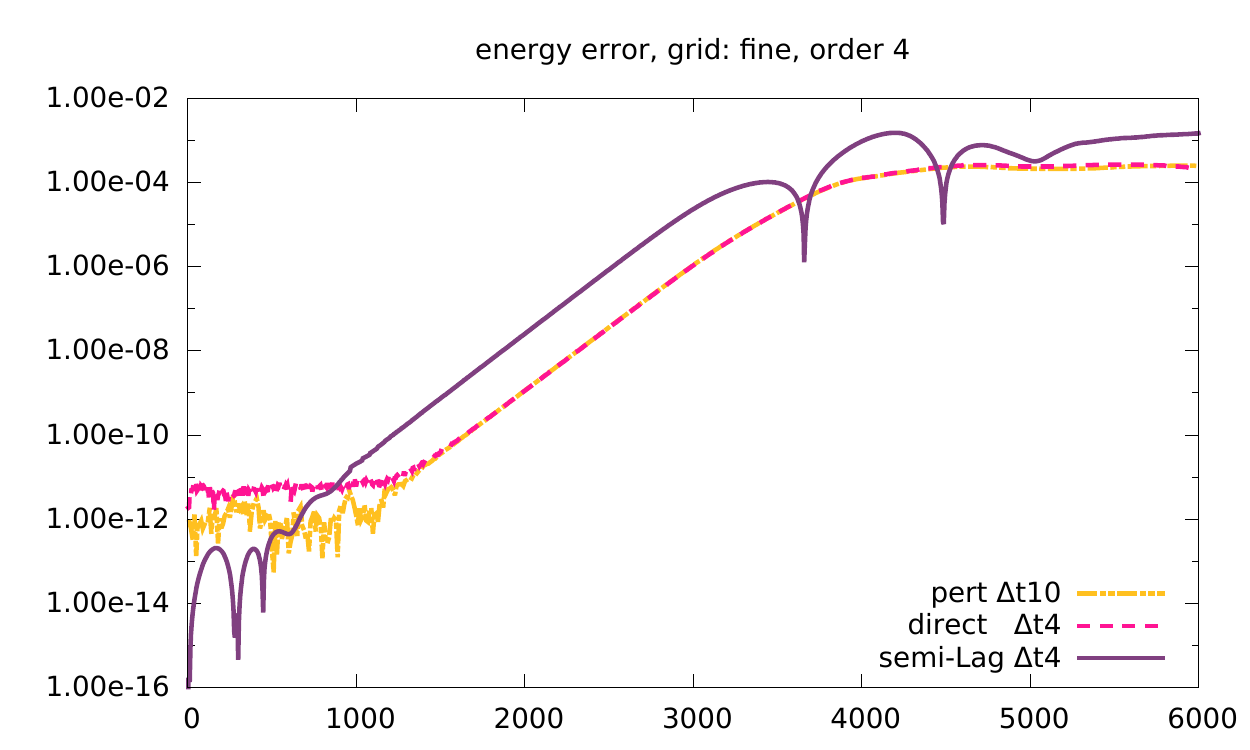}
			\end{tabular}
		\end{center}
		\caption{Comparison of the error in energy for the fourth order exponential integrator and the splitting/semi-Lagrangian method. On the top, the coarse problem which uses $32\times 32\times 32\times 64$ grid points is depicted, while on the bottom the fine grid which uses $64\times 64\times 64\times 128$ points is shown.}
		\label{figenergy_o4}
	\end{figure}

	The $L^2$ error shown in Fig. \ref{figL2_o4} and the energy error shown in Fig. \ref{figenergy_o4} depict similar behavior compared to the second order exponential integrator. In particular, for the $L^2$ error we see a small improvement compared to the semi-Lagrangian approach, while for the energy error we observe an improvement in the error by approximately one order of magnitude.%
	
	\subsection{Mass conservation up to machine precision \label{sec:machineprec}}
	
	As discussed in section \ref{eq:space-discretization} and observed in the previous section, the Arakawa space discretization is not mass conservative up to machine precision (even for homogeneous Dirichlet boundary conditions). Although it should be noted that conservation of mass is still very good (up to $10^{-8}$ in the simulations conducted in the previous section) and, in particular, significantly better than conservation of both momentum and energy. Nevertheless, in some situations it is still beneficial to consider a numerical method that conserves mass up to machine precision. 
	
\begin{figure}
		\begin{center}
			\begin{tabular}{cc}
				\includegraphics[width=0.6\linewidth]{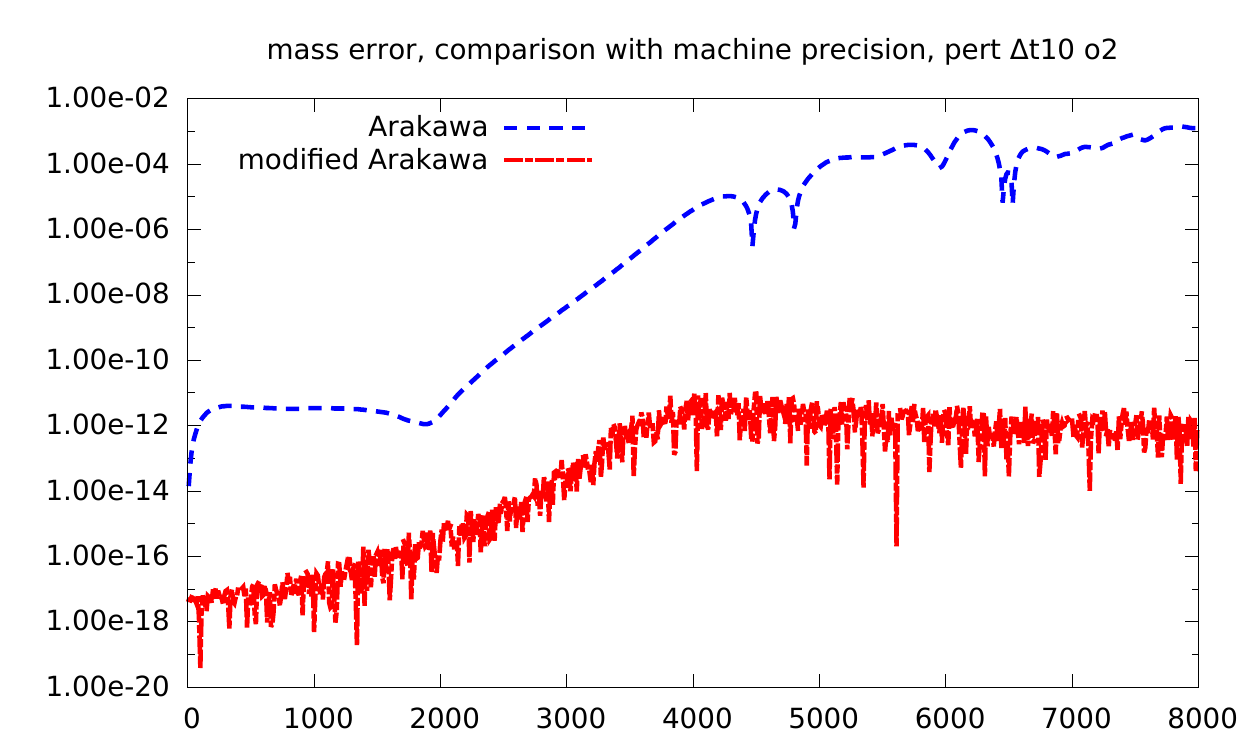} \\
				\includegraphics[width=0.6\linewidth]{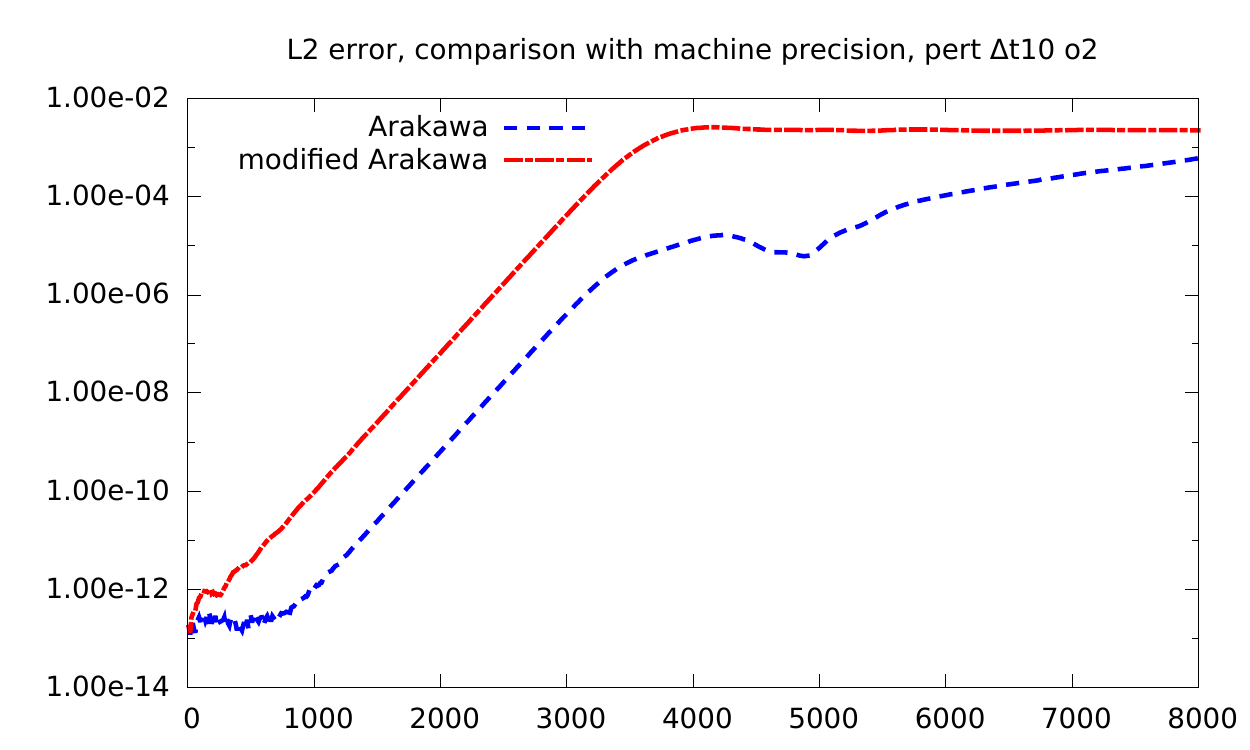} \\
				\includegraphics[width=0.6\linewidth]{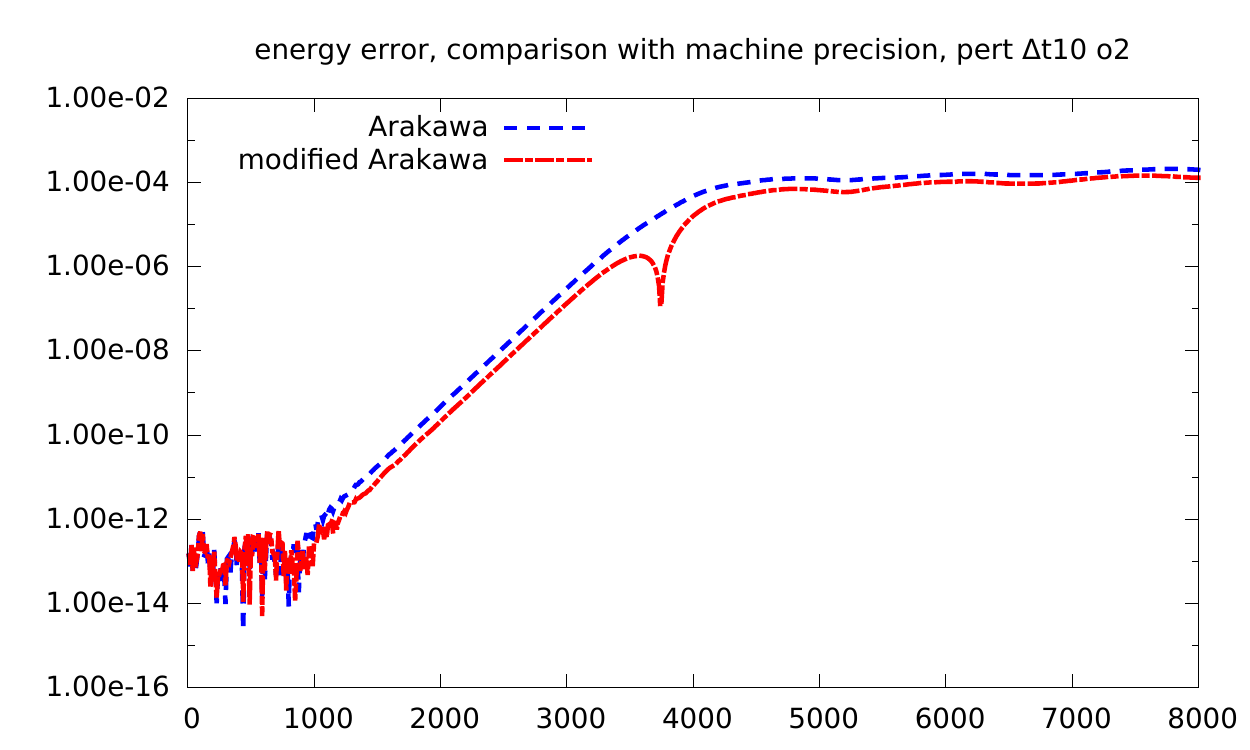}
			\end{tabular}
		\end{center}
		\caption{Comparison of the error in mass (top), $L^2$ norm (middle), and energy (bottom) between the original Arakawa scheme and the modification that preserves mass to machine precision is shown.}
		\label{machine_prec}
	\end{figure}

	Thus, in the following we will consider the performance of the numerical method suggested in section \ref{eq:space-discretization} which conserves mass to machine precision. The numerical results are shown in Fig. \ref{machine_prec}. As expected, we observe conservation of mass up to machine precision which gives a significant advantage compared to standard homogeneous boundary conditions in Arakawa's finite difference method. In addition, we observe a small decrease in the energy error at the expense of a significant increase in the error of the $L^2$ norm. We have also conducted the corresponding simulation with the fourth order exponential integrator. However, since, as expected, the results are almost identical we do not show them here. 
	
	\section{Conclusion \& Outlook\label{sec:conclusion}} %
	
	In this paper we have demonstrated that using exponential integrators is a viable approach for numerically integrating the drift-kinetic equations in time. Combining this with a suitable space discretization (such as Arakawa's method) yields a numerical method with improved conservation properties and significantly reduced computational effort that is still able to take large time steps compared to the most stringent CFL condition. Furthermore, the numerical method proposed in this paper could conceivably be applied to more complicated gyrokinetic models. We consider this as future work.

	In this paper we have exclusively used Fourier techniques in order to solve the linear part of the drift-kinetic equation. However, in principle, any semi-Lagrangian scheme can be used in its place (in the spirit of \cite{latu2}, for example). For example, using a (one-dimensional) cubic spline or discontinuous Galerkin approach would still result in a numerical scheme with mass conservation up to machine precision. In particular the latter, i.e.~using a semi-Lagrangian discontinuous Galerkin approach, might be beneficial in the context of high performance computing systems, where parallelization to a larger number of cores is essential to obtain good performance. We consider this as future work.
	
	\section*{Acknowledgements}
	
	This paper is based upon work supported by the VSC Research Center funded by the Austrian Federal Ministry of Science, Research and Economy (bmwfw). N.C. is supported by the Enabling Research EUROFusion project CfP-WP14-ER- 01/IPP-03 and the IPL FRATRES.

\end{document}